\documentclass[fleqn,usenatbib]{mnras}

\usepackage{newtxtext,newtxmath}

\usepackage{graphicx}    
\usepackage{amsmath}    
\usepackage{bm}
\usepackage{natbib}
\usepackage{color,units}
\usepackage{float}
\usepackage{aas_macros}
\usepackage{hyperref}
\usepackage[T1]{fontenc}
\usepackage{ae,aecompl}

\def\Mtov{M_\text{TOV}}

\newcommand{\citeg}[1]{\citep[e.g.,][]{#1}}

\title[Diversity of magnetar-driven KNe]{On the diversity of magnetar-driven kilonovae}
\author[N.Sarin et al.]{
Nikhil Sarin,$^{1,2}$\thanks{E-mail: nikhil.sarin@su.se}
Conor M. B. Omand,$^{3}$
Ben Margalit$^{4}$\thanks{NASA Einstein Fellow}, 
and David I. Jones$^{5}$
\\
$^{1}$Nordita,
Stockholm University and KTH Royal Institute of Technology
Hannes Alfvéns väg 12, SE-106 91 Stockholm, Sweden\\
$^{2}$The Oskar Klein Centre, Department of Physics, Stockholm University, AlbaNova, SE-106 91 Stockholm, Sweden\\
$^{3}$The Oskar Klein Centre, Department of Astronomy, Stockholm University, AlbaNova, SE-106 91 Stockholm, Sweden\\
$^{4}$Astronomy Department and Theoretical Astrophysics Center, University of California, Berkeley, Berkeley, CA 94720, USA\\
$^{5}$Mathematical Sciences and STAG Research Centre, University of Southampton, Southampton SO17 1BJ, UK}
\date{Accepted XXX. Received YYY; in original form ZZZ}

\pubyear{2022}

\begin{document}
\label{firstpage}
\pagerange{\pageref{firstpage}--\pageref{lastpage}}
\maketitle

\begin{abstract}
A non-negligible fraction of binary neutron star mergers are expected to form long-lived neutron star remnants, dramatically altering the multi-messenger signatures of a merger. Here, we extend existing models for magnetar-driven kilonovae and explore the diversity of kilonovae and kilonova afterglows. Focusing on the role of the (uncertain) magnetic field strength, we study the resulting electromagnetic signatures as a function of the external dipolar and internal toroidal fields. These two parameters govern, respectively, the competition between magnetic-dipole spindown and gravitational-wave spindown (due to magnetic-field deformation) of the rapidly-rotating remnant. We find that even in the parameter space where gravitational-wave emission is dominant, a kilonova with a magnetar central engine will be significantly brighter than one without an engine, as this parameter space is where more of the spin-down luminosity is thermalised. 
In contrast, a system with minimal gravitational-wave emission will produce a kilonova that may be difficult to distinguish from ordinary kilonovae unless early-epoch observations are available. However, as the bulk of the energy in this parameter space goes into accelerating the ejecta, such a system will produce a brighter kilonova afterglow that will peak on shorter times. To effectively hide the presence of the magnetar from the kilonova and kilonova afterglow, the rotational energy inputted into the ejecta must be $\lesssim 10^{-3}-10^{-2} E_{\rm rot}$.
We discuss the different diagnostics available to identify magnetar-driven kilonovae in serendipitous observations and draw parallels to other potential magnetar-driven explosions, such as superluminous supernovae and broad-line supernovae Ic.
\end{abstract}

\section{Introduction}\label{sec:intro}
Multi-wavelength observations of the first binary neutron star merger, GW170817~\citep{abbott17_gw170817_detection, abbott17_gw170817_multimessenger, abbott17_gw170817_gwgrb}, led to an exciting albeit optimistic view of the future of gravitational-wave multi-messenger astronomy. 
Unfortunately, no other gravitational-wave merger has been observed with an electromagnetic counterpart since then. 
However, rapid advances in the capabilities of ground and space-based telescopes in both survey cadence and sensitivity, as well as the first light of upcoming observatories like the Rubin Observatory~\citep{Ivezic2019} and SVOM~\citep{svom}, wide-field telescopes and dedicated gravitational-wave follow-up campaigns~\citeg{evryscope, panstarrs, ztf_paper, goto, kwfi} make it increasingly likely that we will soon start to observe more electromagnetic counterparts to binary neutron star mergers, with and without the gravitational-wave signal.

There are three primary electromagnetic counterparts to a binary neutron star merger observable within seconds to days after the merger itself. 1) The short gamma-ray burst produced by a relativistic jet launched immediately after the merger~\citeg{Eichler:1989Natur, Narayan:1992ApJL}. 2) The kilonova produced by r-process nucleosynthesis in the neutron-rich ejecta~\citeg{li98,metzger10}. 3) The jet afterglow produced by the relativistic jet interacting with the ambient interstellar medium~\citeg{sari98, sari99}. On much longer timescales, there is an additional electromagnetic counterpart, the kilonova afterglow, an analogue to the jet afterglow, but instead produced when the kilonova ejecta interacts with the ambient interstellar medium~\citeg{Nakar2011}. There are tentative hints that we may have now begun to see the kilonova afterglow of GW170817~\citep{Hajela2021}. 
The properties of all counterparts are connected to the progenitor, the aftermath of the merger, the environment and the observer's viewing angle. However, many of these connections, particularly the links to the progenitor properties, are not well understood. We refer the reader to~\citet{Nakar2020} for a detailed recent review of the electromagnetic counterparts to a binary neutron star merger (see also \citealt{Metzger&Berger12}).

As traditionally classified, there are four possible outcomes for a binary neutron star merger. These depend on the mass of the remnant and the nuclear equation of state, which sets the maximum allowed non-rotating neutron star mass, $\Mtov{}$~\citeg{Bernuzzi2020, Sarin2021_review}. For remnant mass $M \gtrsim 1.2 \Mtov{}$, a black hole is born either promptly after the merger or over thermal/viscous timescales. Alternatively, for $M \lesssim 1.2 \Mtov{}$, a meta-stable neutron star can be born in the aftermath of the merger, and can survive for up to $\unit[10^{4}]{s}$~\citep{ravi14} or indefinitely if $M \lesssim \Mtov{}$
(however see \citealt{Margalit+22} for a recent exploration of the post-merger remnant fate that challenges this traditional classification scheme). 

The fate of GW170817 is not definitively known, but the bulk of evidence from the inferred energetics of the afterglow and kilonova suggests that the remnant of GW170817 was a \textit{hypermassive} neutron star that collapsed into a black hole within $\mathcal{O}(\unit[1]{s})$~\citeg{Margalit2017, ruiz18_170817, Gill2019}. Although, GW170817 observations are consistent with a \textit{supramassive} neutron star that survived for up to $\approx \unit[3000]{s}$ if the remnant spins down predominantly through gravitational-wave radiation~\citep{ai20}. 
The only other observed binary neutron star merger, GW190425, was not observed with an electromagnetic counterpart but likely promptly collapsed into a black hole given its high total mass~\citep{abbott20_gw190425}. 
Observations of double neutron star systems in our Galaxy and beyond~\citeg{Galaudage2021}, population synthesis studies~\citep{Chattopadhyay:2019xye, Broekgaarden2021}, and inferences into $\Mtov{}$~\citeg{Margalit2017, ruiz18_170817, shibata19, Sarin2020a, Raaijmakers2021} suggest that a significant fraction of binary neutron star mergers will produce a long-lived neutron star remnant~\citep{Margalit2019, Sarin2022}. 

A long-lived neutron star remnant provides the system with an extensive reservoir of rotational energy that can dramatically alter the timescales and energetics of the aforementioned electromagnetic counterparts. This rotational energy reservoir has been used to explain many different features of various electromagnetic transients. Such as the X-ray plateaus of gamma-ray bursts~\citeg{rowlinson10,Dall'Osso2011,rowlinson13,lasky17,Strang2019, Sarin2020b, Strang2021}, some fast X-ray transients~\citep{Xue2019, Xiao2019, Yang2019, Ai2021, Lin2022}, or bright kilonovae~\citeg{fong20_kilonova}. 
However, in many cases, the same observations could be interpreted with the afterglow produced by a structured relativistic jet~\citep{Beniamini2020, Sarin2021_cdf}, or attributed to systematics and not require a neutron star central engine~\citep{zhu20, oconnor21}. 
Theoretically, the impact of the long-lived neutron star remnants, in particular on kilonovae, has been explored in detail in many works~\citeg{Yu2013, Metzger2014a, Metzger2014b, Siegel&Ciolfi16, li18, Wollaeger2019, Kawaguchi2020, Kawaguchi2021, Kawaguchi2022, Ai2022}. However, these works either do not explore the entire parameter space or make different simplifying assumptions in modelling, such as ignoring relativistic dynamics, gamma-ray leakage, pair cascades, the dynamical evolution of the neutron star, or gravitational-wave emission. 

This paper explores the diversity of engine-driven kilonovae, relaxing some of the simplifying assumptions made in previous works. In Sec.~\ref{sec:model}, we introduce our model for the magnetar spin evolution and the engine-driven kilonova model. In Sec~\ref{sec:results}, we explore the diversity of kilonovae and kilonova afterglow signatures due to differences in the magnetic field strength of the nascent neutron star. In Sec.~\ref{sec:diagnostics}, we discuss the perils of interpreting observations of engine-driven kilonovae with models without an engine, kilonova modelling systematics, and discuss diagnostics that could be used to infer the presence of a long-lived neutron star engine. 
We discuss the implications of jets, other gravitational-wave emission mechanisms, orthogonalisation timescales, the magnetar wind nebula spectrum, draw parallels to other potential engine-driven transients and conclude in Sec.~\ref{sec:implications}. Throughout the paper, we use the notation $Q_x = Q/10^x$ in cgs units unless otherwise noted.
\section{Model}\label{sec:model}
As mentioned above, the rotational energy of a long-lived remnant neutron star provides a large additional energy reservoir $\sim\unit[10^{53}]{erg}$ cf. $\sim\unit[10^{51}]{erg}$ for an ordinary kilonova \citep{Metzger2015_transientdiversity,Margalit2019}. This additional energy can radically change the signature of the kilonova and kilonova afterglow. 
However, not all of this reservoir is available to increase the brightness of the kilonova, as some energy may be lost through gravitational-wave emission, or not thermalised, or potentially lost in launching a relativistic jet, or lost in the collapse of the neutron star to a black hole, etc. We discuss the case of a remnant that collapses in more detail in Sec.~\ref{sec:implications}. 
To accurately predict the light curves of a magnetar-driven kilonova, one needs to consider the initial conditions of the neutron star established shortly after merger, such as the spin period, magnetic field strengths, and radius. These initial conditions directly affect the dynamical evolution of the nascent neutron star and, therefore, dictate how much of the rotational energy reservoir is available to increase the kilonova's luminosity. 

A neutron star post-merger remnant depletes its rotational energy through a combination of electromagnetic and gravitational-wave radiation i.e., 
\begin{equation}
    - \frac{dE_{\rm rot}}{dt} = L_{\rm EM} + L_{\rm GW}.
    \label{eqn:derotdt}
\end{equation}
Here $L_{\rm EM}$ is the energy lost through electromagnetic radiation, which assuming vacuum dipole radiation is~\citep{Spitovsky2006}, 
\begin{equation}
    L_{\rm EM} = \frac{\mu^2\Omega^4}{4c^3}(1 + \sin^2 \chi), 
    \label{eqn:L_em}
\end{equation}
where $\mu = B_{\rm ext}R_{\rm NS}^3$ is the dipole moment of the neutron star, $B_{\rm ext}$ is the surface magnetic dipole field strength, $R_{\rm {NS}}$ and $\Omega$ are the radius and angular rotational frequency of the neutron star, respectively, and $\chi$ is the angle between the magnetic and rotation axes. 
The assumption of vacuum dipole radiation is likely incorrect; almost no pulsars observed in our Galaxy have a braking index consistent with vacuum dipole radiation~\citep{archibald16, Lower2021}.
Additionally, recent modeling of {\it Neutron Star Interior Composition Explorer} (NICER) observations favor complex multipolar external magnetic fields over simple dipole configurations \citep{Bilous+19,Miller+19,Riley+19}. 
Furthermore, simulations of realistic pulsar braking indices are also inconsistent with that expected from magnetic dipole radiation~\citep{melatos97, Bucciantini2006}. Moreover, many measurements of braking indices of putative nascent neutron stars born in gamma-ray bursts are inconsistent with vacuum dipole radiation~\citep{lasky17, Xiao2019_braking, Lu2019}. However, the differences due to assuming vacuum dipole radiation as opposed to a more realistic electromagnetic torque are small and unnecessary for studying the diversity of magnetar-driven kilonovae as intended in this work.  
The above expression likely only holds after $\approx \unit[40]{s}$~\citep{Lander2020}, before which the $L_{\rm EM}$ depends on the wind mass-loss rate and magnetisation~\citep{Metzger2011}; our simulations will not attempt to capture this very early phase of the spin-down. We note that a small $\mathcal{O}(\unit[10^{51}]{erg})$ of rotational energy can also be lost through neutrino emission~\citeg{shibata19}, which we ignore here.

The second term in Eq.~\ref{eqn:derotdt} represents the gravitational-wave luminosity emitted by a rapidly-rotating deformed (asymmetric) neutron star. This can be expressed as \citep{Cutler00}
\begin{equation}
    L_{\rm GW} = \frac{2}{5}\frac{G(I \epsilon_{\rm B})^2}{c^5} \Omega^6 \sin^2 \chi (1 + 15\sin^2 \chi). 
    \label{eqn:L_gw}
\end{equation}
Here $I$ and $\epsilon_{\rm B}$ are the moment of inertia and ellipticity of the neutron star. 
Under the assumption that the ellipticity is generated by an internal toroidal magnetic field, we can relate the ellipticity to the strength of the field via
\begin{equation}
    \epsilon_{\rm B} = -3.0 \times 10^{-4} \left(\frac{B_{\rm int}}{B_{\rm ext}}\right)^2 B_{\rm ext, 16}^2,
    \label{eqn:ellip}
\end{equation}
where $B_{\rm int}$ is the volume averaged internal toroidal field~\citep{Cutler2002}.  The negative sign indicates that the field distorts the star into a prolate shape. An internal poloidal magnetic field would tend to produce a deformation of opposite sign (oblate).  The toroidal field/prolate deformation case is the relevant one to us, as such systems naturally evolve to the orthogonal rotator configuration ($\chi = \pi/2$) considered here \citep{Cutler00, Lander2018, Lander2020}.

We note that the expression for $L_{\rm{GW}}$ and $\epsilon_{\rm{B}}$ above do not capture gravitational-wave radiation mechanisms such as a bar mode instability or r-modes which could be active in these newly born systems~\citep{corsi09, andersson01}. 
However whether they are active on relevant timescales and grow to amplitudes large enough to make them relevant to our discussion here is not well understood~\citeg{andersson03, doneva15, lasky16}.

The expressions for both $L_{\rm EM}$ and $L_{\rm GW}$ indicate two critical aspects. 1) The increased sensitivity of the gravitational-wave luminosity to $\chi$, i.e., the angle between the magnetic field and rotation axes. 2) The increased sensitivity of the gravitational-wave luminosity to the angular velocity of the neutron star, $\Omega^{6}$ cf. $\Omega^{4}$ for electromagnetic radiation. 
The former makes it essential to consider realistic values of $\chi$ or, ideally, a reliable prescription for its evolution, while the increased sensitivity to $\Omega$ makes the choice of initial conditions like the initial spin period fundamental.  
For the spin period, we note that numerical simulations suggest that the remnant is likely to be rotating near the mass-shedding limit (i.e., $p_{0} \lesssim \unit[1]{ms}$)~\citeg{Bernuzzi2020}. 
However, this is sensitive to when the gravitational-wave losses from the viscous phase cease~\citep{Radice2018}. 
Constraints of the initial spin period from putative neutron stars born in short gamma-ray bursts suggest $p_{0} \lesssim \unit[5]{ms}$~\citep{rowlinson13}.  

With the above ingredients in hand, we can now model how the star's rotational energy reservoir interacts with the kilonova ejecta and affects its dynamics and luminosity. Our model is similar in respect to the one-dimensional "merger-nova'' model presented in~\citep{Yu2013}. However, we make some modifications to account for energy losses due to pair cascades~\citep{Metzger2014a}, and model the efficiency of converting spin-down energy into internal energy of the ejecta as a time-varying quantity. In particular, we assume the efficiency is dictated by the gamma-ray leakage of the ejecta~\citep{Wang2015}, similarly to models of neutron star engines applied to superluminous supernovae~\citeg{Nicholl2017}.

The ultra-relativistic magnetar wind expands and collides with the expanding ejecta, decelerating and pushing a forward shock through the ejecta within seconds~\citep{Gao2013}. This wind pushes on and accelerates the ejecta, increasing its kinetic energy and internal energy. The total energy of the ejecta can be expressed as the combination of its kinetic and internal energy, 
\begin{equation}
    E_{\rm ej} = (\Gamma - 1)M_{\rm ej}c^2 + \Gamma E'_{\rm int}.
    \label{eq:eej}
\end{equation}
Here $\Gamma$ is the ejecta Lorentz factor, $M_{\rm ej}$ is the ejecta mass, and $E'_{\rm int}$ is the internal energy in the co-moving rest frame.  
The evolution of this system depends on the interplay between the energy sources i.e., the radioactive heating and magnetar spin-down luminosity and the energy loss channels i.e., the radiated luminosity and the adiabatic expansion of the ejecta. The dynamical evolution of the ejecta is therefore, 
\begin{equation}
    \frac{d\Gamma}{dt} = \frac{\xi L_{\rm EM} + L_{\rm ra} - L_{\rm bol} - \Gamma \mathcal{D}(dE'_{int}/dt')}{M_{\rm ej}c^2 + E'_{\rm int}}.
    \label{eq:dgamdt}
\end{equation}
Here $L_{\rm ra}$ and $L_{\rm bol}$ are the radioactive power and emitted bolometric luminosity respectively, $\xi$ is the fraction of electromagnetic spin-down luminosity injected into the ejecta, $\mathcal{D} = 1/[\Gamma(1 - \beta)]$ is the relativistic Doppler factor with $\beta = \sqrt{1-\Gamma^{-2}}$, and $dt'$ is the co-moving time which can be connected to the observer time by $dt' = \mathcal{D}dt$. 
In previous models derived in the literature~\citep{Yu2013, Metzger2019, Ai2022}, $\xi$ is assumed to be some constant and decoupled from the properties and evolution of the ejecta. Here, we relax this assumption, modelling $\xi$ to vary with time coupled to the gamma-ray leakage of the ejecta. In particular, we model $\xi$ as 
\begin{equation}
   \xi = 1 - e^{-At^{-2}},
   \label{eq:xi}
\end{equation}
where
\begin{equation}
    A = \frac{3 \kappa_\gamma M_{\rm ej}}{4\pi v^2_{\rm ej}},
    \label{eq:leakage}
\end{equation}
is the leakage parameter~\citep{Wang2015} and $\kappa_\gamma$ is the gamma-ray opacity of the ejecta. 

The last term in the numerator in Eq.~\ref{eq:dgamdt} describes the evolution of the internal energy of the ejecta, and can be written as~\citep{Kasen2016} 
\begin{equation}
  \frac{dE'_{\rm int}}{dt'} = \xi L'_{\rm EM} + L'_{\rm ra} - L'_{\rm bol} - \mathcal{P}'\frac{dV'}{dt'}.
  \label{eq:deint_dt}
\end{equation}
Here the first two terms on the right-hand side capture the energy gained from the spin-down luminosity of the nascent neutron star and radioactive heating, while the third and fourth terms capture the energy emitted away from the system and lost due to expansion of the ejecta, respectively. We note that the prime indicates quantities in the co-moving rest frame, which can be related to the relevant unprimed quantities via $L'_x = L_x/\mathcal{D}^2$. 

The radioactive power in the co-moving frame is given by 
\begin{equation}
    L'_{\rm ra} = 4 \times 10^{49}  M_{\rm ej, -2} \left[ \frac{1}{2} - \frac{1}{\pi}\arctan\left(\frac{t' - t'_o}{t'_{\sigma}} \right)\right]^{1.3} \text{ erg s$^{-1}$},
    \label{eq:l_ra}
\end{equation}
with $t'_o \sim 1.3$ s and $t'_{\sigma} \sim 0.11$ s~\citep{Korobkin2012}. In principle, $L'_{\rm{ra}}$ may also have an efficiency term due to neutrino or gamma-ray leakage which we ignore here for simplicity. 

In a kilonova with a magnetar engine, $\xi L'_{\rm EM}$ can dwarf the radioactive power $L'_{\rm ra}$ and the total energy available from the magnetar is significantly larger than the initial kinetic energy of the ejecta. Therefore, models of magnetar-driven kilonovae need to consider the work done by the expansion of the ejecta. The work done by free ejecta expansion $\mathcal{P}dV'$ converts internal energy into bulk kinetic energy.  The pressure $\mathcal{P} = E'_{\rm int}/3V'$ is dominated by radiation, and the evolution of the co-moving volume $V'$ is 
\begin{equation}
    \frac{dV'}{dt'} = 4\pi R^2_{\rm ej} \beta c,
    \label{eq:dvdt}
\end{equation}
where the evolution of the ejecta radius $R_{\rm ej}$ is
\begin{equation}
    \frac{dR_{\rm ej}}{dt} = \frac{\beta c}{1 - \beta}.
    \label{eq:drdt}
\end{equation}

The radiated bolometric luminosity can be derived approximately from the diffusion equation in the co-moving frame~\citep{Kasen2010, Kotera2013}

\begin{align}
    L'_{\rm bol} = & \frac{E'_{\rm int}c}{\tau R_{\rm ej}/\Gamma} = \frac{E'_{\rm int}t'}{t'^2_{\rm diff}}, & \text{ for $t \leq t_\tau$}, \label{lbol_pret} \\
    = & \frac{E'_{\rm int}c}{R_{\rm ej}/\Gamma}, & \text{ for $t > t_\tau$}, \label{lbol_postt}.
\end{align}
Here, 
\begin{equation}
    \tau = \frac{\kappa M_{\rm ej} R_{\rm ej}}{V'\Gamma}
    \label{eq:opdep}
\end{equation}
is the optical depth, $\kappa$ is the ejecta opacity, 
\begin{equation}
    t'_{\rm diff} = \left(\frac{\tau R_{\rm ej} t'}{\Gamma c}\right)^{1/2}
    \label{eq:tdiff}
\end{equation}
is the effective diffusion time, and $t_\tau$ is the time when $\tau = 1$ which is always greater than $t_{\rm diff}$.

To extract the radiated bolometric luminosity of a magnetar-driven kilonova, we numerically solve Eqs.~\ref{eq:dgamdt} and~\ref{eq:deint_dt} governing the dynamical evolution of the system and the internal energy of the ejecta, respectively, using Eqs.~\ref{eqn:derotdt}-\ref{eqn:L_gw} for the magnetar spin evolution, and Eqs.~\ref{eq:dvdt} and~\ref{eq:drdt} for the evolution of the ejecta. 
We then also account for suppression of the observed luminosity due to pair cascades~\citep{Metzger2014a,Kasen2016,Metzger2019} via 
\begin{equation}
    L_{\rm obs} = \frac{L_{\rm bol}}{1 + (t_{\rm life}/t)}.
    \label{eq:pairobs}
\end{equation}
Here
\begin{align}
    \frac{t_{\rm life}}{t} = & \frac{\tau_n v}{c(1-\alpha)} , \\
    \approx & \frac{0.6}{1 - \alpha} Y_{-1}^{1/2} L_{\rm EM, 45}^{1/2} \left(\frac{v}{0.3c}\right)^{1/2} \left(\frac{t}{1 \text{ day}}\right)^{-1/2}
    \label{eq:tlife}
\end{align}
is the characteristic lifetime of a nebular non-thermal photon compared to the dynamical timescale, $\alpha$ is the frequency-averaged albedo of the ejecta, $Y$ is the fraction of electromagnetic spin-down power $L_{\rm EM}$ that is converted into electron/positron pairs, and $\tau_n \approx 900$ s is the neutron half-life. 

We can then use the idealisation of a blackbody spectrum to calculate the effective temperature as 
\begin{equation}
    T_{\rm eff} = \left(\frac{L_{\rm obs}}{4\pi\sigma_{\rm SB} R^2_{\rm ph}}\right)^{1/4},
\end{equation}
where $\sigma_{\rm SB}$ is the Stefan-Boltzmann constant and $R_{\rm ph}$ is the radius of the photosphere. The flux density at frequency $\nu$ is then 
\begin{equation}
    F_{\nu}(t)=\frac{2 \pi h \nu^{3}}{c^{2}\mathcal{D}(t)^2} \frac{1}{\exp \left[h \nu / \mathcal{D}(t) k_{\rm B} T_{\mathrm{eff}}(t)\right]-1} \frac{R_{\mathrm{ph}}^{2}(t)}{D_{L}^{2}},
\end{equation}
where $k_{\rm B}$ is the Boltzmann constant and $D_{L}$ is the luminosity distance to the source. 
\section{Diversity}\label{sec:results}
We now explore the diversity of magnetar-driven kilonovae and the kilonova afterglows they produce for different initial conditions using the model derived in Sec.~\ref{sec:model}. In particular we first explore the impact of the internal and external magnetic field strengths of an infinitely stable neutron star with radius $R_{\rm NS} = \unit[11]{km}$, moment of inertia $I = \unit[3 \times 10^{45}]{g~cm^{2}}$, initial spin period of $p_0 = \unit[0.7]{ms}$ i.e., a neutron star rotating at approximately the mass-shedding limit unless otherwise stated, the frequency average albedo of the ejecta $\alpha=0.5$, and $Y = 0.05$, i.e., the fraction of the neutron star electromagnetic spin-down luminosity that is lost due to pair cascade emission. 
We also fix the ejecta mass $M_{\rm ej} = 0.05 M_{\odot}$, typical of what might be expected with a neutron star remnant \citep{Margalit2019} and consistent with observations of AT2017gfo~\citeg{Villar2017}; initial $v_{\rm ej} = \unit[0.2]{c}$; $\kappa = 1$ cm$^2$ g$^{-1}$, which is typical of lanthanide-poor ejecta~\citep{Metzger2019}; and $\kappa_\gamma = 0.1$ cm$^2$ g$^{-1}$. 

We note that we do not explore the variation from different values of ejecta mass and velocity. In general, larger initial ejecta kinetic energies minimise the differences between a magnetar-driven kilonova and one without. By contrast, low ejecta kinetic energies make the differences between an engine-driven and no-engine kilonovae more profound. However, the broad features of magnetar-driven kilonovae and the diversity due to differences in magnetic field strengths remain the same.

The value of $\kappa$ can be larger by $\sim$an order of magnitude if the ejecta is rich in lanthanides ($\kappa \sim 10 {\rm cm}^2 \, {\rm g}^{-1}$; \citealt{Barnes&Kasen13,Kasen+13,Tanaka&Hotokezaka13,Tanaka2020}). However in our context, neutrino irradiation by the long-lived merger remnant is expected to raise the electron fraction of surrounding material, leading to lanthanide-poor ejecta whose opacity is lower (e.g., \citealt{Metzger2014b}).
Our choice of $\kappa_\gamma$ is consistent with the range found by \citealt{Hotokezaka2020}. However, we note that~\citet{Hotokezaka2020} considered the opacity to gamma-rays from radioactive decay, which need not be the same as $\kappa_\gamma$ for nebular gamma-rays that is relevant in our context.
The value of $\kappa_\gamma$ is also motivated by the gamma-ray opacity inferred from modeling magnetar-powered SLSNe~\citep{Nicholl2017} and is broadly consistent with results from detailed calculations of nebular gamma-ray escape~\citep{Vurm&Metzger21}. However, the different ejecta environments between magnetar-driven kilonovae and SLSNe may lead to different $\kappa_\gamma$, even if the magnetar-wind nebula itself is similar. 
In general, the time at which $\xi$ begins to drop below $\approx 1$ (i.e., the time where the choice of $\kappa_\gamma$ matters) is roughly $t_\gamma \equiv (3\kappa_\gamma M_{\rm ej}/4\pi v_{\rm ej}^2)^{1/2}$. This timescale is related to the time of peak optical emission $t_{\rm pk}$ as $t_\gamma = t_{\rm pk} (c/v_{\rm ej})^{1/2} (\kappa_\gamma/\kappa)^{1/2}$. In our kilonova context $c/v_{\rm ej} \lesssim 5$ and $\kappa_\gamma/\kappa \sim 0.1$ so $t_\gamma \lesssim t_{\rm pk}$ and gamma-ray leakage starts shortly before peak emission. By contrast, for typical SLSNe $c/v_{\rm ej} \gtrsim 10$ and $\kappa_\gamma/\kappa \sim 1$ so the efficiency $\xi$ drops below $\approx 1$ only after the optical light-curve peaks. This implies that the choice of $\kappa_\gamma$ in kilonovae can be a source of systematic uncertainty for observations near and after the optical peak. However, the broad features and diversity in magnetar-driven kilonovae due to differences in magnetic field strengths remain the same for all choices of $\kappa_\gamma$.
We note that high values of $\kappa_\gamma$ drive $\xi$ towards $\approx 1$, consistent with efficiency values used in magnetar-driven models in the literature previously~\citep{Yu2013, Metzger2019}. 

The assumption of an infinitely stable remnant provides the system with up to $E_{\rm rot} \approx 6 \times 10^{52} P^2_{-3}$ erg of rotational energy available to affect the dynamics and luminosity of the kilonova or emitted in gravitational waves.
We further assume the neutron star is an orthogonal rotator, i.e., $\chi = \pi/2$, consistent with detailed numerical simulations that find that neutron stars are driven towards this state for the first $\unit[10^6]{s}$ of their lifetimes~\citep{Lander2020}. 
The impact of the orthogonalisation timescale, evolution of $\chi$, other channels of energy losses such as powering a jet, the collapse of the neutron star into a black hole, and other gravitational-wave emission mechanisms are discussed in more detail in Sec.~\ref{sec:implications}. 
\subsection{Kilonova Energy Budget}
Only the energy available in electromagnetic radiation (ignoring the energy lost in potentially powering a jet) can impact the kilonova and kilonova afterglow. Here, we explore what fraction $f_{\rm EM}$ of $E_{\rm rot}$ is available for different internal and external magnetic field strengths; i.e., what fraction of the rotational energy is emitted as electromagnetic radiation as opposed to gravitational waves following the evolution of a neutron star using Eq.~\ref{eqn:derotdt}. 

\begin{figure}
    \centering
    \includegraphics[width=\columnwidth]{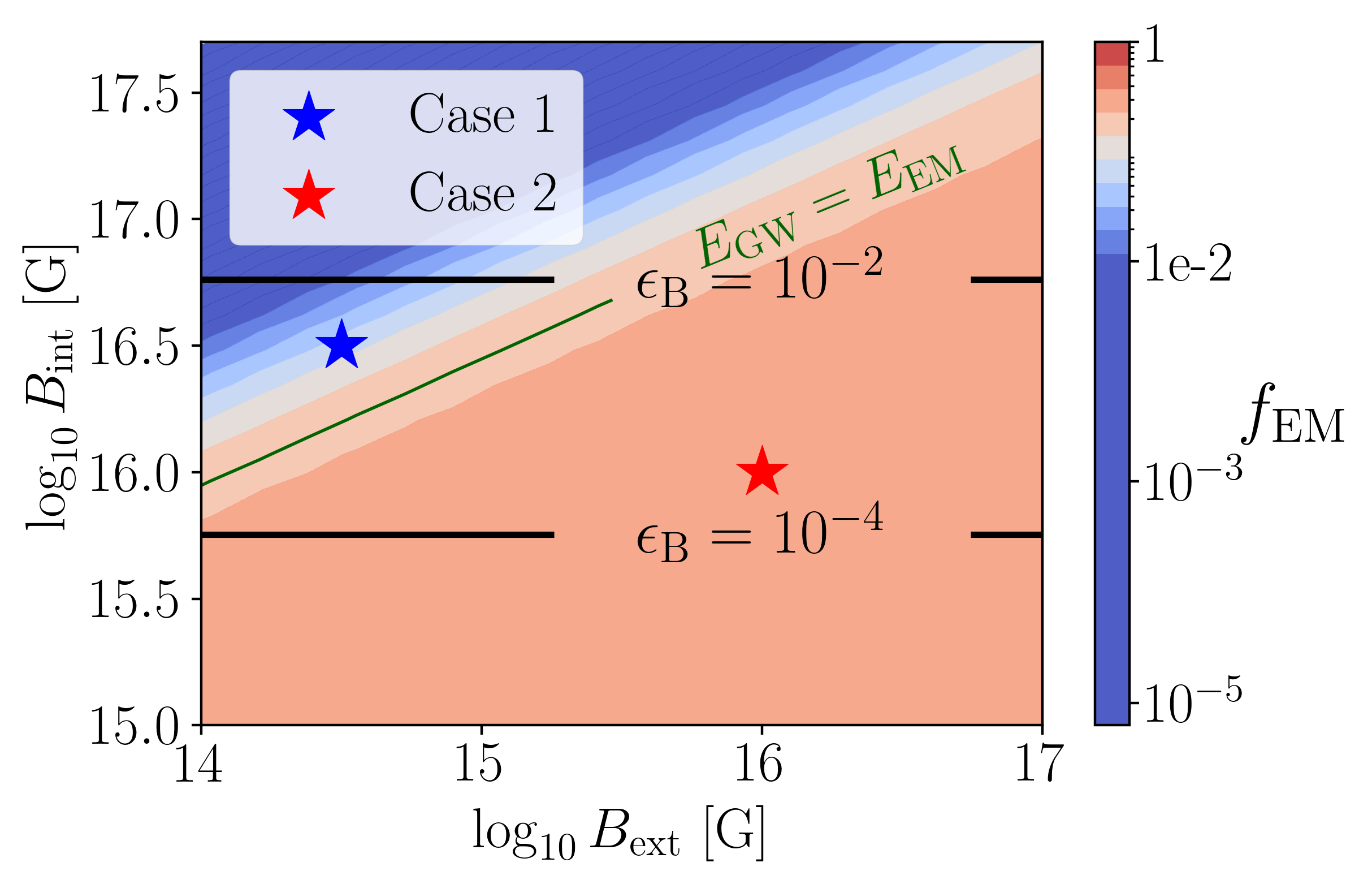}
    \caption{Fraction of spin-down energy $E_{\rm rot}$ emitted as electromagnetic radiation for various strengths of external dipole field $B_{\rm ext}$ internal toroidal field $B_{\rm int}$. For a neutron star with radius $R_{\rm NS} = \unit[11]{km}$, moment of inertia $I = \unit[3 \times 10^{45}]{g~cm^{2}}$, and initial spin period of $p_0 = \unit[0.7]{ms}$. The black lines show where $\epsilon_{\rm B} = 10^{-2}$, which is an upper limit on what could be considered physically reasonable magnetic deformation induced ellipticity, and $\epsilon_{\rm B} = 10^{-4}$. The green curve shows the magnetic field strengths where the emitted electromagnetic and gravitational wave energy is equal. The stars show the parameters for two representative cases which we explore in more detail in Sec.~\ref{sec:cases}.}
    \label{fig:f_EM}
\end{figure}

Figure~\ref{fig:f_EM} shows the fraction of the total cumulative rotational energy $f_{\rm EM}$ that is radiated electromagnetically (rather than gravitationally), and is therefore available to change the kilonova and kilonova afterglow signature. In particular, we evolve Eq.~\ref{eqn:derotdt} with parameters described above for a grid of magnetic field strengths over the first $\unit[1000]{days}$ of the remnant's lifetime. Note that the remnant loses the bulk of its rotational energy on the spin-down timescale, which is significantly shorter than $\unit[1000]{days}$ for the entirety of this parameter space. We evaluate $f_{\rm EM}$ for grids on the internal and external magnetic fields from $\unit[10^{14}-10^{18}]{G}$. However, given magnetic flux conservation and magnetic-field amplification processes such as the magnetorotational and Kelvin-Helmholtz instabilities during the merger process, we expect most neutron stars born in binary neutron star mergers to have external magnetic fields $B_{\rm ext} \gtrsim \unit[10^{15}]{G}$. The strength of the internal toroidal field is difficult to predict; differential rotation in the early phase of the remnant's lifetime is expected to generate a significant internal toroidal field, something seen in numerical simulations~\citep{Rezzolla2011, Kiuchi2014, Moesta+20}. However, whether a particular field configuration is stable is unclear. The parameters used in Fig~\ref{fig:f_EM} demonstrate the most optimistic (pessimistic) scenario for gravitational-wave (electromagnetic) radiation. Higher initial spin periods or a different angle between the spin and magnetic field axes would reduce the parameter space where gravitational-wave radiation (at least from a magnetic deformation) are relevant.

Across the bulk of the parameter space, the majority of the rotational energy is available as electromagnetic radiation i.e., $f_{\rm EM} \gtrsim 0.6$. Gravitational waves losses only become significant if the internal toroidal field is approximately two orders of magnitude larger than the external poloidal field, which only appears realistic if $B_{\rm ext} \lesssim \unit[10^{15}]{G}$.
This is broadly consistent with previous results by \cite{Margalit2017} (see their Fig 3).
Although we note that the required difference becomes smaller as the overall magnetic field strength increases. For example, for a neutron star with $B_{\rm ext}= \unit[10^{14}]{G}$ to emit equal amounts of gravitational-wave and electromagnetic radiation requires an internal toroidal field of $B_{\rm int} \sim \unit[10^{16}]{G}$. However, a neutron star with an external field $B_{\rm ext}= \unit[10^{16}]{G}$, will emit equivalent amount of gravitational-wave and electromagnetic radiation for an internal toroidal field $B_{\rm int} \sim \unit[10^{16.5}]{G}$. 
A large portion of the parameter space where gravitational-wave radiation is dominant has a magnetic-deformation induced neutron star ellipticity $\epsilon_{\rm B} \gtrsim 10^{-2}$, requiring uncomfortably large internal toroidal fields. However, such low fractions of $f_{\rm EM}$ could be achieved due to other gravitational-wave emission mechanisms such as the bar-mode or r-mode instability, which will also deplete the rotational energy reservoir. 
We note that even in the highly optimistic case that the entire rotational energy budget is radiated in gravitational waves, a gravitational-wave signal would likely still be undetectable with advanced LIGO at design sensitivity out to distances beyond $\sim \unit[20]{Mpc}$~\citep{Gao2017, Sarin2018, Sur2021}. 
\subsection{Kilonova Energetics and Timescales}
Aside from losses due to gravitational waves, other inefficiencies stop us from using the magnetar spin-down luminosity as a direct proxy for the overall increase in luminosity of the kilonova. 1) Only some fraction (encapsulated by $\xi$) of the magnetar electromagnetic spin-down luminosity gets coupled to the evolution of the ejecta. 
2) Not all coupled energy is thermalised into radiation. 
In particular, depending on the ratio of the diffusion and spin-down timescales, the bulk of the energy accelerates the ejecta or gets thermalised into radiation. The former is particularly relevant, as changes to the kinetic energy of the ejecta can radically alter the kilonova afterglow brightness and peak timescales. 

\begin{figure}
    \centering
    \includegraphics[width=\columnwidth]{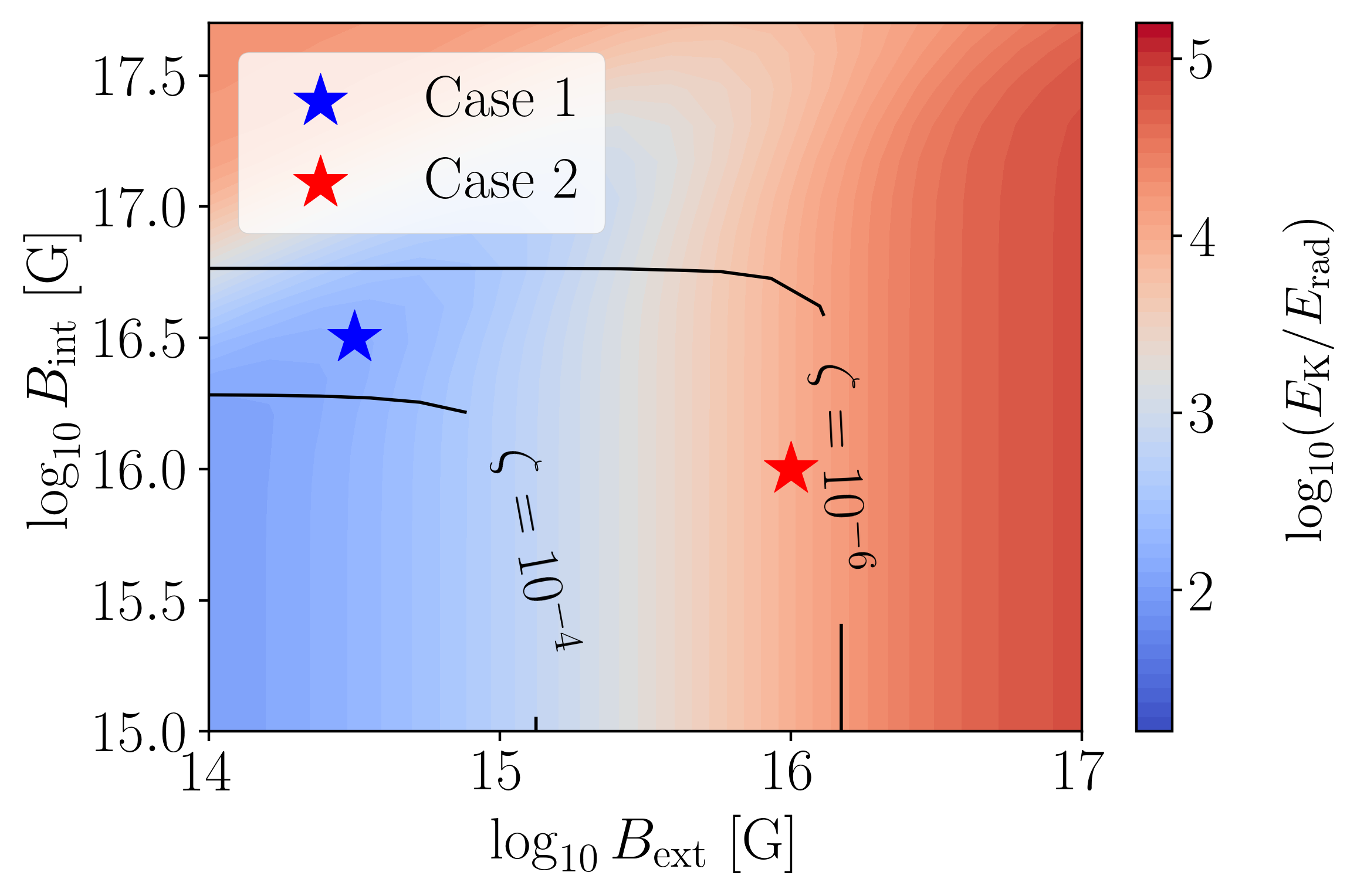}
    \caption{Ratio of ejecta kinetic energy $E_{\rm K}$ to total radiated energy $E_{\rm rad}$ for various strengths of external dipole field $B_{\rm ext}$ internal toroidal field $B_{\rm int}$. Across the entire parameter range $E_{\rm K} \gg$ $E_{\rm rad}$. The black curves show where $\zeta = t_{\rm SD}/t_{\rm diff}$ is 10$^{-4}$ and 10$^{-6}$. The stars show the parameters for two representative cases which we explore in more detail in Sec.~\ref{sec:cases}.}
    \label{fig:calor}
\end{figure}

We first explore what fraction of $f_{\rm EM}$ is converted into kinetic energy and into radiation. In Fig~\ref{fig:calor}, we show the ratio of the kinetic energy to the radiated energy for the grid of magnetic field strengths as above. Two features of the figure are immediately noticeable. 1) across the entire parameter space, more energy is converted into kinetic energy than is radiated, i.e., most of the neutron star's electromagnetic spin-down luminosity accelerates the ejecta instead of increasing the kilonova luminosity. 
For the most energetic neutron star engines, this translates into an absolute upper limit on the amount of radiated energy in a magnetar-driven kilonova; no magnetar-driven kilonova will have $E_{\rm rad} \geq 4 \times \unit[10^{51}]{erg}$. 2) Above $B_{\rm int} \sim \unit[10^{16.7}]{G}$, the ratio of the kinetic and radiated energy becomes insensitive to the magnetic field strength. This is approximately the internal field where $\epsilon_B = 10^{-2}$ and is unlikely to be achieved. We also show contours for $\zeta = t_{\rm SD}/t_{\rm diff}$, i.e., the ratio of the spin-down and diffusion timescale; the former set by the remnant and the latter set by initial properties of the ejecta and the amount of kinetic energy imparted to it by the remnant. Higher values of $\zeta$ imply higher values of radiated energy, while smaller values imply more energy lost in accelerating the ejecta. This is consistent with numerical results found for magnetar-driven supernovae~\citep{Suzuki2021}.

\begin{figure}
    \centering
    \includegraphics[width=\columnwidth]{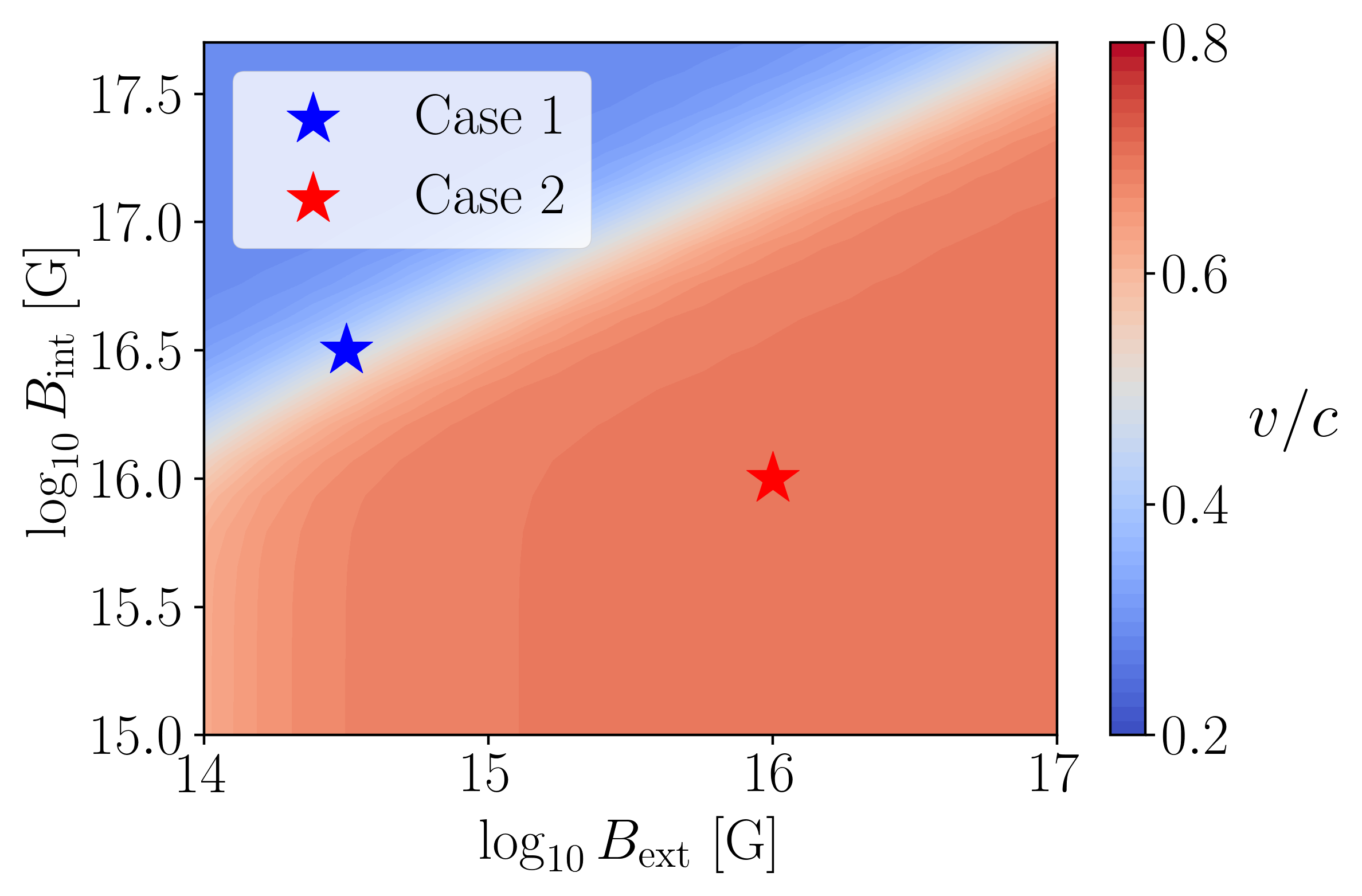}
    \caption{Ejecta velocity for various strengths of external dipole field $B_{\rm ext}$ internal toroidal field $B_{\rm int}$. The stars show the parameters for two representative cases which we explore in more detail in Sec.~\ref{sec:cases}.}
    \label{fig:gamma}
\end{figure}

We next examine the bulk velocity of the kilonova ejecta across the magnetic field strength parameter space, which we show in Figure~\ref{fig:gamma}. 
The ejecta is accelerated from the initial velocity $\unit[0.2]{c}$ to $\sim \unit[0.8]{c}$ ($\Gamma \sim$ 1.4) over all of the parameter space where electromagnetic emission dominates. Interestingly, the curve which separates the region with substantial acceleration over the initial value is slightly above the curve corresponding to $E_{\rm GW} = E_{\rm EM}$ shown in Fig~\ref{fig:f_EM}. 
This is consistent with physical intuition, as the ejecta can only be significantly accelerated when there is sufficient electromagnetic energy inputted to change the dynamics of the ejecta set by the initial conditions. This highlights the importance of considering the dynamical evolution of the ejecta when there is a central engine present. 
We note that the plot above corresponds to velocity for an ejecta mass $M_{\rm ej} = \unit[0.05]{M_{\odot}}$, smaller values of ejecta mass will be accelerated to much higher velocities and vice versa.
This is particularly important as a true kilonova will likely have a distribution of velocities or at least two or more components~\citeg{Metzger2019}. For example, for ejecta masses $\lesssim 0.01 M_{\odot}$, the material can reach $\Gamma \sim 10$ in certain parts of the parameter space. Such high Lorentz factor ejecta can have a substantial impact on the relativistic jet afterglow. We discuss the effect of different ejecta properties in more detail in Sec.~\ref{sec:implications}.

\begin{figure}
    \centering
    \includegraphics[width=\columnwidth]{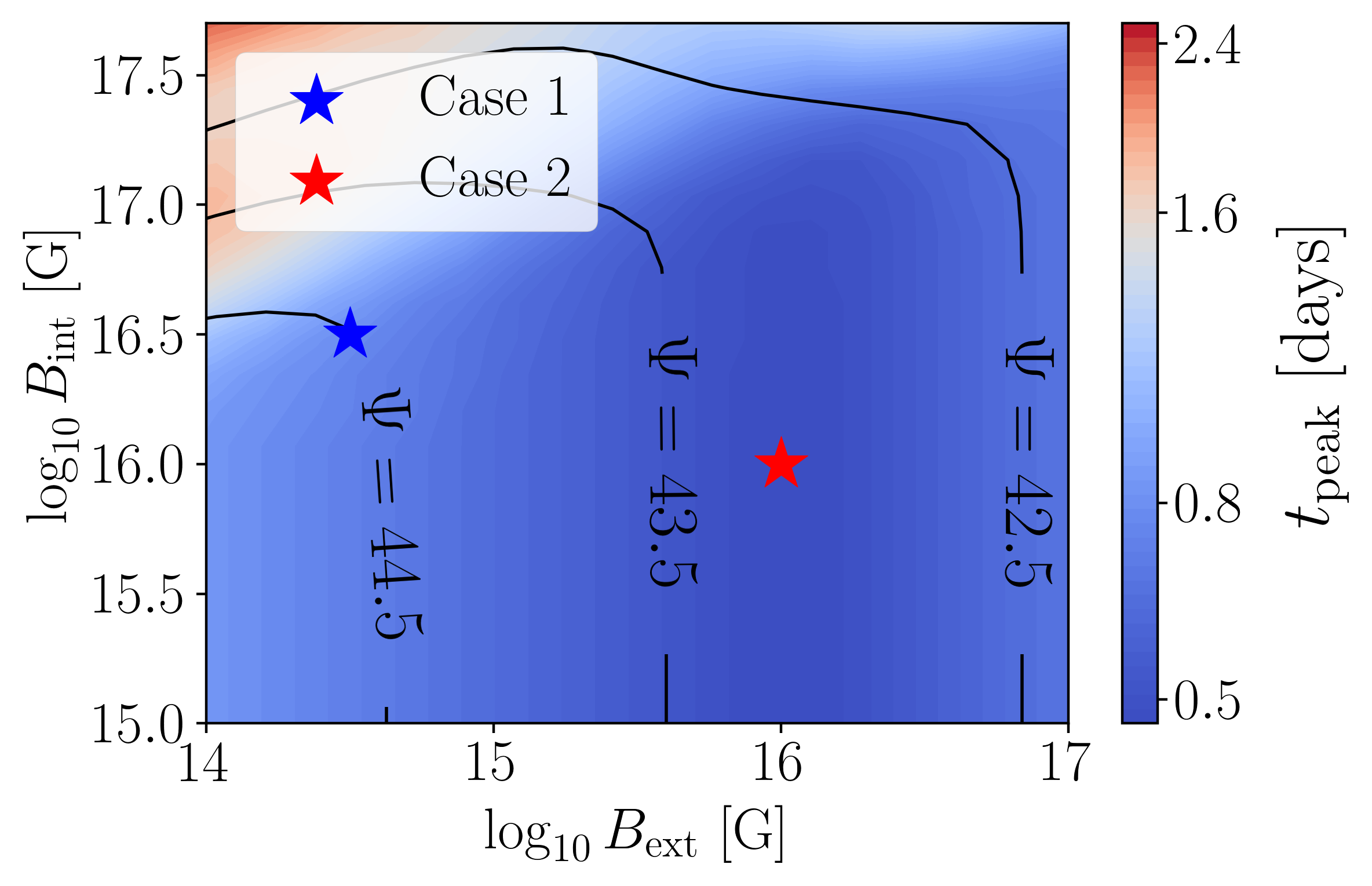}
    \caption{Bolometric peak timescale for the kilonova for different strengths of external dipole field $B_{\rm ext}$ internal toroidal field $B_{\rm int}$. The black curves show $\Psi = \log_{10} (L_{\rm {bol, peak}}~[\rm{erg}])$ i.e., the peak bolometric luminosity. The stars show the parameters for two representative cases which we explore in more detail in Sec.~\ref{sec:cases}.}
    \label{fig:peaktime}
\end{figure}

Perhaps the most critical aspect of discovering magnetar-driven kilonovae is to understand their peak timescales. Kilonovae are fast transients, and magnetar-driven kilonovae are faster still, stressing the need for surveys with high cadence. 
In Fig~\ref{fig:peaktime} we show the kilonova (bolometric) peak timescale for the same grid of magnetic field strengths as in previous figures.
There is minimal dependence in the parameter space when there is substantial electromagnetic radiation energy in the system, i.e., high $f_{\rm EM}$. 
This is consistent with physical intuition as for all regions of parameter space where $f_{\rm} \sim 1$ the engine has already depleted almost all of its rotational energy reservoir before the light curve peaks. 
In this region, the peak time varies between $\unit[0.4-1.6]{days}$, while the region where gravitational-wave radiation dominates the kilonova peak timescale is $\approx \unit[1.6-2.4]{days}$, consistent with the peak timescale of a kilonova without an engine with $M_{\rm ej} = 0.05 M_{\odot}$, $v_{\rm ej} = \unit[0.4]{c}$, and $\kappa = 1$ cm$^2$ g$^{-1}$~\citep{Metzger2019}. 
The first optical detection of AT2017gfo was at $\sim$ $\unit[11]{hours}$~\citep{abbott17_gw170817_multimessenger}, comparable or later than the peak time for cases where the neutron star remnant has altered the kilonova dynamics. 
However, as we stress in Sec.~\ref{sec:diagnostics}, it becomes increasingly difficult to distinguish a magnetar-driven kilonova from an ordinary one after the kilonova peaks making it imperative that potential candidates are rapidly followed up. 

\begin{figure}
    \centering
    \includegraphics[width=\columnwidth]{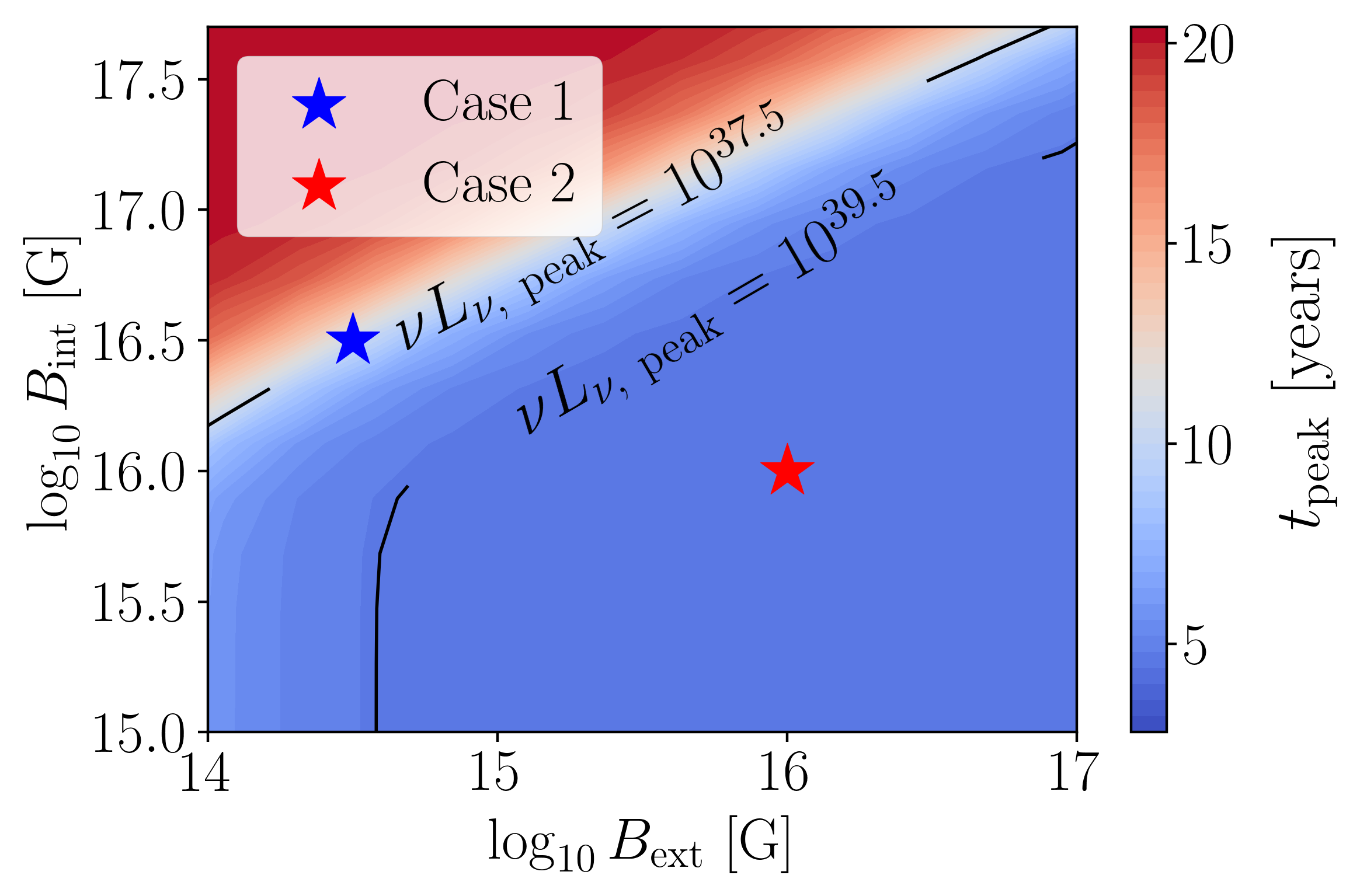}
    \caption{X-ray ($\unit[1]{keV}$) kilonova afterglow peak timescale for various strengths of external dipole field $B_{\rm ext}$ internal toroidal field $B_{\rm int}$. The black curves show the peak $\nu L_{\nu}$ X-ray afterglow luminosity. The stars show the parameters for two representative cases which we explore in more detail in Sec.~\ref{sec:cases}.}
    \label{fig:afterglowpeaktime}
\end{figure}

We also explore how the presence of the neutron star impacts the peak timescale of the kilonova synchrotron afterglow \citep{Nakar2011}. In particular, we calculate the synchrotron afterglow produced across the grid of magnetic field strengths above following~\citet{schroeder20}.
We assume a fiducial  interstellar medium density of $10^{-2}$ cm$^{-3}$, that the fraction of energy in non-thermal electrons and shock-amplified magnetic fields are $0.1$ and $0.01$ respectively, and an electron power-law slope of $2.5$. These values are guided by analyses of short gamma-ray burst afterglows~\citeg{fong15}. We then calculate the $\nu L_\nu$ peak timescale of the $\unit[1]{keV}$ X-ray afterglow, shown in Figure \ref{fig:afterglowpeaktime}.
While quantitative values of these properties depend sensitively on the assumed interstellar density and microphysical parameters, the trends with $B_{\rm int}$ and $B_{\rm ext}$ are insensitive to these choices. 
We note that for simplicity we have here neglected the contribution of thermal electrons to the afterglow signature, though this can potentially be important given the mildly-relativistic ejecta velocities~\citep{Margalit2021}. 
We also neglect the affect of a jet on the ejecta afterglow (see \citealt{Margalit2020}).

The kilonova afterglow peaks on the deceleration timescale of the ejecta~\citep{Nakar2011}. 
In Fig.~\ref{fig:afterglowpeaktime}, this kilonova afterglow peak time demonstrates similar features to the kilonova peak timescale, with some variation across the parameter space where electromagnetic emission dominates.
This is consistent with physical intuition as the only parameter that changes across this parameter space is the velocity and, therefore, the kinetic energy of the ejecta (see Fig.~\ref{fig:calor} and~\ref{fig:gamma}), with faster ejecta velocity corresponding to earlier peak times. 
Notably, the kilonova afterglow peak timescale becomes longer in the parameter space where gravitational-wave radiation dominates. In particular, the curve representing a peak timescale of $\sim \unit[10]{years}$ is consistent with the curve representing equal amounts of energy lost in gravitational-wave radiation and available as electromagnetic radiation seen in Fig.~\ref{fig:f_EM}. 
This is consistent with intuition, as in this parameter space the bulk of the rotational energy is lost in gravitational-wave radiation, and, therefore, the peak timescale is similar to the peak timescale for a system without an engine. 
\subsection{Case Studies: Bright, Fast, and Kilonovae without an engine} \label{sec:cases}
With the impact of the magnetic field strengths explored, we now examine three representative cases in more detail to illustrate the impact of other parameters and show representative lightcurves. To wit, ``Case 1'' with $B_{\rm int} = \unit[10^{16.5}]{G}$ and $B_{\rm ext} = \unit[10^{14.5}]{G}$, ``Case 2'' with $B_{\rm int} = \unit[10^{16}]{G}$ and $B_{\rm ext} = \unit[10^{16}]{G}$ and, for comparison purposes, a kilonova without an engine. 
All other parameters are the same as in the previous section unless specified. The two cases represent two intriguing parts of the parameter space; ``Case 1'' represents a system where the nascent neutron star loses a significant amount of energy through gravitational-wave radiation, but the larger fraction of the available electromagnetic radiation is thermalised, producing a more luminous kilonova. Alternatively, ``Case 2'' represents a scenario where almost all the rotational energy is available as electromagnetic radiation, but the bulk of the energy goes into accelerating the ejecta instead of being thermalised. The two cases represent a kilonova that is ``brighter'' or ``faster'' than an ordinary kilonova without an engine, respectively. However, we caution that this description does not hold across all of the parameter space. 

\begin{figure}
    \centering
    \includegraphics[width=\columnwidth]{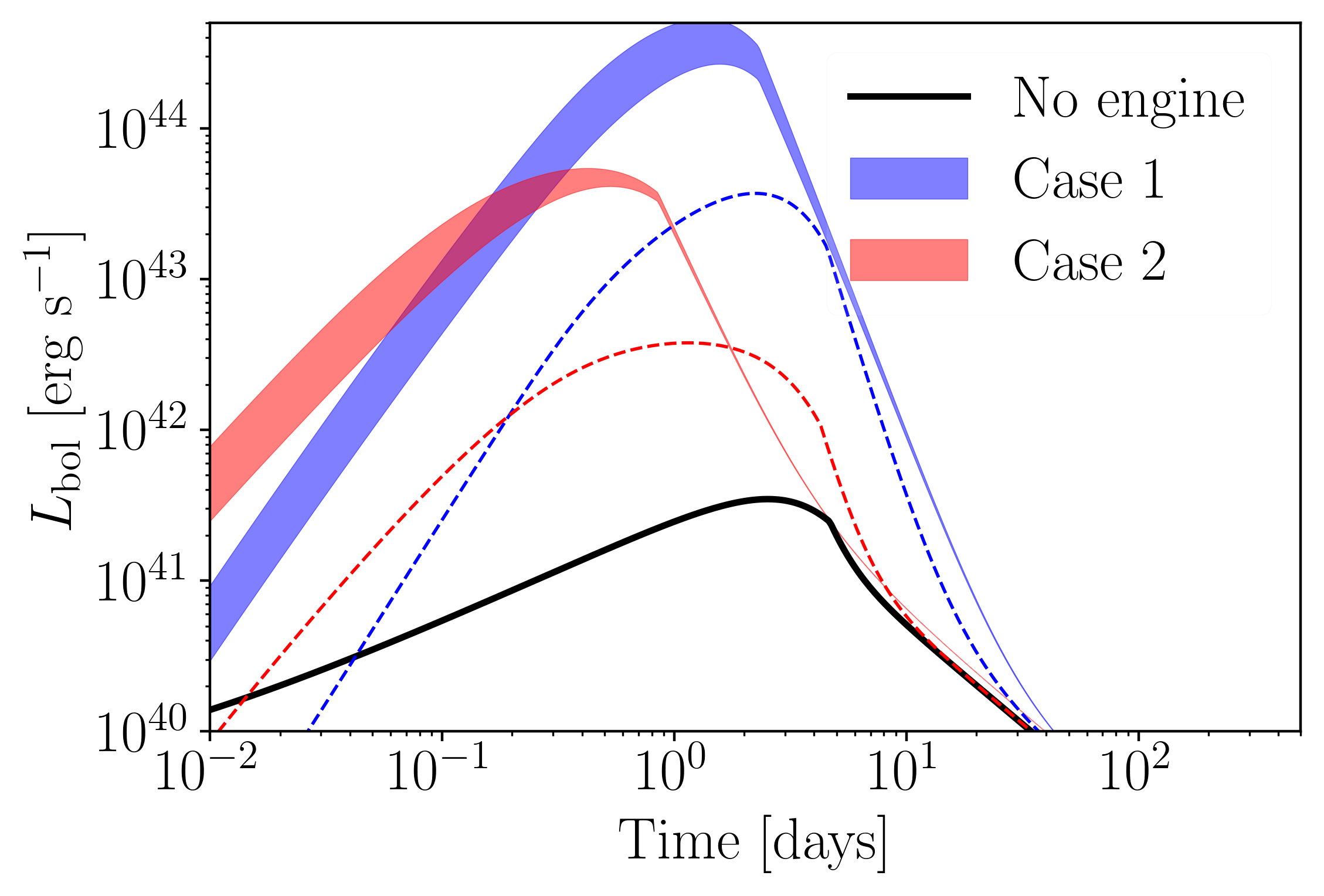}
    \caption{Bolometric light curves for the three representative kilonovae; ``Case 1'', ``Case 2'', and a kilonova without an engine in blue, red, and black, respectively. The coloured band in the two magnetar-driven kilonovae represent the uncertainty due to the unknown fraction of neutron star spin-down luminosity converted into pair cascades ($Y$ from 0.01 to 0.1 (see Equation \ref{eq:tlife}). Dashed curves show light curves for the same representative cases but for a different initial spin period $p_0 = \unit[5.0]{ms}$.}
    \label{fig:lcs}
\end{figure}

In Fig.~\ref{fig:lcs}, we show the bolometric light curves for these representative kilonovae. ``Case 1'', ``Case 2'' and a kilonova without an engine are shown in blue, red, and black, respectively, with the dashed curves representing the bolometric light curves for the same parameters but an initial spin period $p_0 = \unit[5]{ms}$. 
The bands in the blue and red curves represent the uncertainty due to the unknown fraction of neutron star electromagnetic spin-down luminosity lost due to pair cascade emission, i.e., we vary the parameter $Y$ (see Equation \ref{eq:tlife}) between 0.01-0.1 to cover the range of plausible values~\citep{Svensson1987, Lightman1987, Metzger2019}. 

Consistent with intuition, the peak bolometric luminosity of the two magnetar-driven kilonovae ($\sim 10^{43-45}$ erg s$^{-1}$) is larger than the kilonova without an engine with a peak luminosity of $\sim$ 10$^{41}$ erg s$^{-1}$. 
In particular, ``Case 1'' is brighter than ``Case 2'' despite the former losing significant energy in gravitational waves. Again, this is consistent, as a larger fraction of the available energy is thermalised in ``Case 1'' compared to ``Case 2''. 
The peak timescales also differ for all models, with ``Case 2'' kilonova peaking earliest at $\sim \unit[0.7]{days}$, ``Case 1'' peaking at $\sim \unit[1]{day}$, similar to the peak timescale of the kilonova without an engine.
The ``Case 2'' kilonova becomes hard to distinguish from a kilonova without an engine after the latter peaks at $\sim \unit[2]{day}$, while the ``Case 1'' kilonova remains significantly brighter than the kilonova without an engine for up to $\sim \unit[30]{days}$. 
The effect of varying $Y$ (see Equation \ref{eq:tlife}) can change the luminosity of the light curve by a factor of at most $\sim 2-3$.
However, this does not meaningfully affect the interpretation of the light curve. 
By contrast, a change in $p_0$ can create more than an order of magnitude difference (shown by the dashed curves), which could affect the light curve interpretation. 
These results stress the importance of high cadence surveys and rapid follow-up, especially to distinguish between a ``Case 2''-type magnetar-driven kilonova and a kilonova without an engine. 
We note that other systematics with brightness related diagnostics suggest that this alone is likely, not conclusive evidence for a magnetar-driven kilonova, and further diagnostics such as the colour evolution and spectra may be better proxies~\citep{Metzger2014b, Metzger2019}. We discuss these systematics and diagnostics to infer the presence of a magnetar in kilonovae in more detail in Sec.~\ref{sec:diagnostics}. 

\begin{figure*}
    \centering
    \includegraphics[width=0.8\paperwidth]{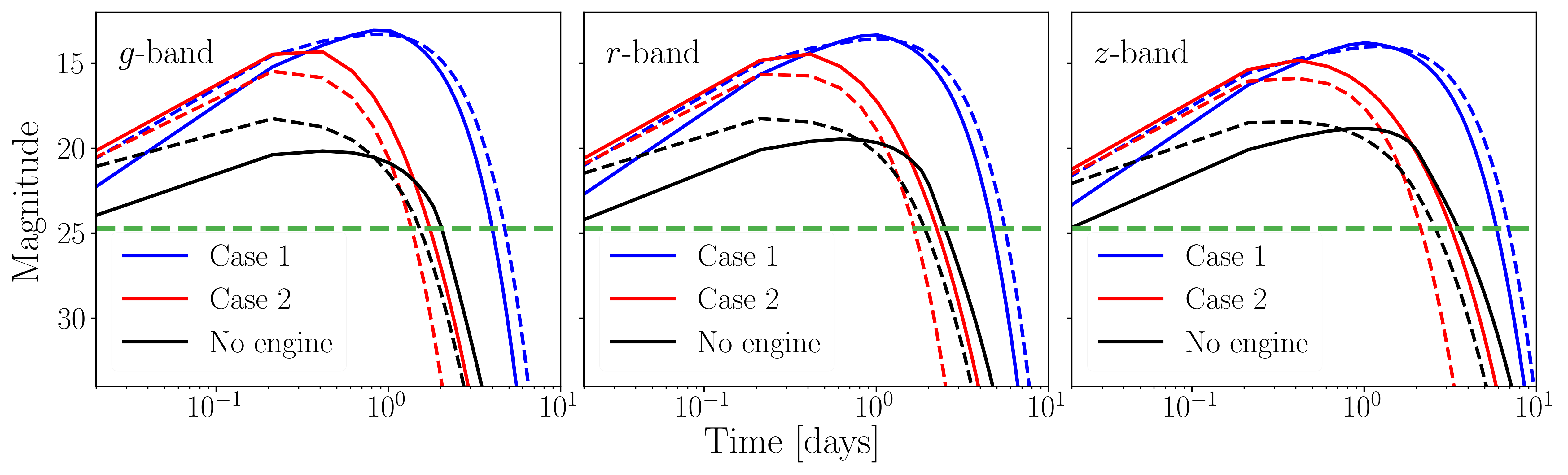}
    \caption{The $g$, $r$, and $z$-band light curves for the three representative kilonovae; ``Case 1'', ``Case 2'', and a kilonova without an engine in blue, red, and black, respectively at $\unit[100]{Mpc}$. The green horizontal band indicates the Vera Rubin detection threshold ($\unit[24.7]{Mag}$). The dashed blue, red, and black curves are kilonova lightcurves for the same three representative cases but using a model motivated by~\citet{Metzger2019}.}
    \label{fig:photometry}
\end{figure*}

In Fig.~\ref{fig:photometry}, we show the corresponding $g$, $r$, $z$-band photometry for these three kilonovae at a distance of $\unit[100]{Mpc}$. The dashed green lines indicate the magnitude limit of Vera Rubin~\citep{Ivezic2019}; for comparison, a current-generation optical telescope, such as the Zwicky Transient Facility (ZTF), has an $r$ and $g$ band magnitude limit of $20.5$. We note that ZTF does not have a $z$-band filter. 
The dashed curves represent the same three representative kilonovae but modelled using the kilonova model from~\citet{Metzger2019} with the same magnetar evolution and physics of gamma-ray leakage as described in Sec.~\ref{sec:model} to illustrate the systematics between different kilonova models. 
We discuss these systematics and differences in models in more detail in Sec.~\ref{sec:diagnostics}.
For both kilonova models and all three representative cases, there is no discernible difference between the $g$ and $r$-band photometry, i.e., the $g$ and $r$-band photometry for a particular representative kilonova and model show minimal variation. In particular, all three cases peak at a consistent magnitude and peak and become too dim for Vera Rubin at a consistent time. However, the $z$-band photometry shows significant differences, peaking earlier and becoming too dim for Vera Rubin at least a day later.

For both the model derived in Sec.~\ref{sec:model} (solid curves) and the model from~\citet{Metzger2019} (dashed curves), ``Case 1'' is the brightest kilonova and is also detectable for the longest time, consistent with the bolometric luminosity shown in Fig.~\ref{fig:lcs}. 
There is also a minimal difference between the solid and dashed blue curves (especially after peak), suggesting that the systematic uncertainty from different kilonova models is minimal for this representative kilonova. 
The same can not be said for the other two representative cases. For the same model, ``Case 2'' is brighter than a kilonova without an engine at peak. However, there is a significant difference in the peak magnitude, peak time, and overall evolution across the two models. 
In particular, ``Case 2'' peaks at the same time for both models at a magnitude $\sim 15$, with the~\citet{Metzger2019} model always dimmer by a few magnitudes.
However, the model motivated by~\citet{Metzger2019} predicts a brighter ``no engine'' kilonova that also stays detectable for $\sim \unit[1]{day}$ longer in Vera Rubin compared to the ``Case 2'' kilonova. 
The model derived in Sec.~\ref{sec:model} is not as bright without an engine, with a difference of $\approx 7$ magnitudes compared to the ``Case 2'' kilonova when the latter peaks cf. a difference $\approx 3$ magnitudes for the model motivated by~\citet{Metzger2019}. The difference in brightness is even more pronounced before peak, with the ``no engine'' and ``Case 2'' kilonova having an $\approx 2$ magnitude difference at $\unit[0.1]{days}$ cf. $\approx 6$ magnitude difference between the two representative cases for the model derived in Sec.~\ref{sec:model}. 

The largest difference in brightness across both models and representative cases occurs when the engine-driven kilonova peaks. For a ``Case 1''-like kilonova and the parameters described above the peak is at $\sim \unit[1-2]{days}$ in the $g$, $r$, and $z$-band compared to a ``Case 2''-like kilonova which peaks at $\unit[0.1-0.3]{days}$. The latter timescale is significantly shorter than the $\unit[11]{hours}$ it took for the first observations of AT2017gfo~\citep{abbott17_gw170817_multimessenger}, stressing the need for ultra-high cadence surveys e.g., Tomo-e Gozen~\citep{Sako2016, Sako2018, Kojima2018} for serendipitous observations and low-latency or negative latency alerts for gravitational-wave follow-up.

\begin{figure*}
    \centering
    \includegraphics[width=0.8\paperwidth]{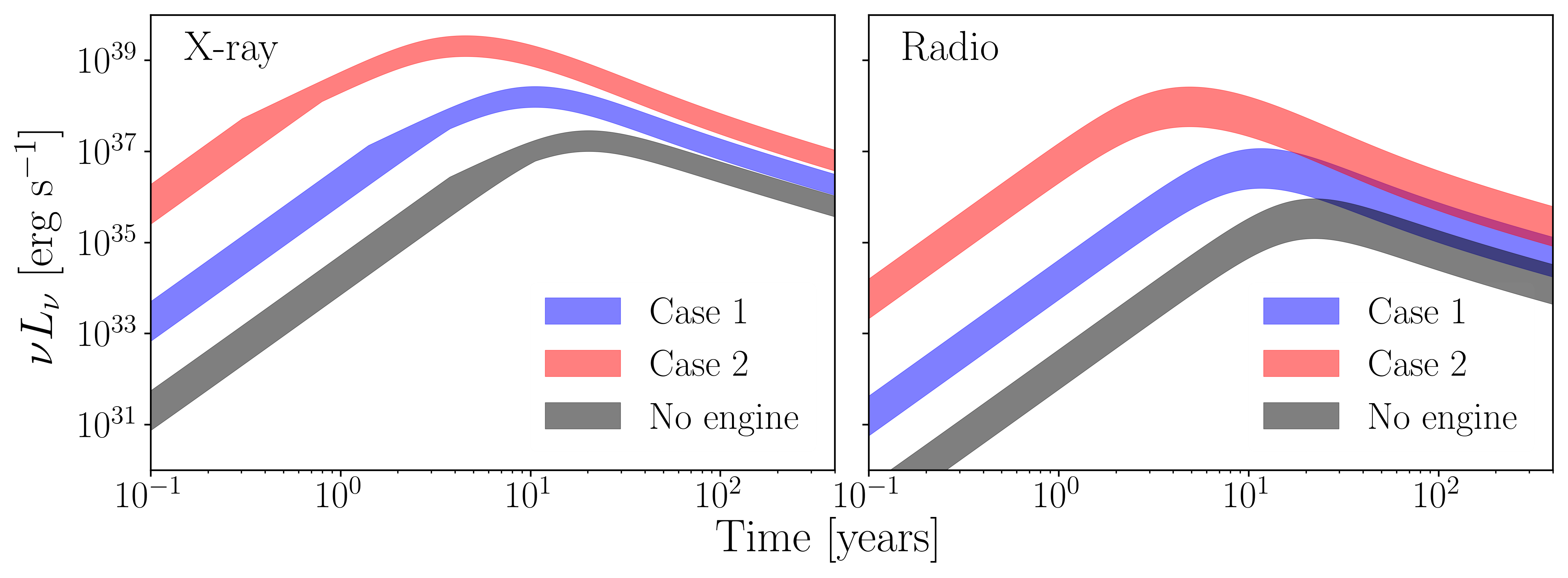}
    \caption{X-ray (1 keV) and radio (1 GHz) kilonova afterglow light curves on the left and right panels for the three representative kilonovae; ``Case 1'', ``Case 2'', and a kilonova without an engine in blue, red, and black, respectively. The shaded band shows the variability from an order of magnitude uncertainty on the interstellar medium density ($\unit[10^{-3}-{10^{-2}}]{cm^{-3}}$).}
    \label{fig:afterglowlcs}
\end{figure*}

We now turn to explore the kilonova afterglow for the three representative cases. In Fig.~\ref{fig:afterglowlcs}, we show the x-ray (1 keV) and radio (1 GHz) kilonova afterglows from the three models in the left and right panels, respectively. 
We use the same fiducial parameters as described previously, with the shaded band indicating the uncertainty due to the unknown interstellar medium density, which we vary from $\unit[10^{-3}-10^{-2}]{cm^{-3}}$, consistent with constraints on GRB170817A~\citeg{Hajela+19}. 

The kilonova afterglow illustrates three salient aspects: 1) A ``Case 2''-like kilonova ends up becoming more easily distinguishable from the kilonova afterglow without an engine ($\nu L_{\nu} \approx \unit[10^{39}]{erg~s^{-1}}$ at peak cf. $\approx \unit[10^{37}]{erg~s^{-1}}$ for a `Case 1''-like or no-engine system) i.e., the kilonova afterglow brightness is a better diagnostic for a ``Case 2''-like system in contrast to a ``Case 1''-like system where the kilonova itself is a better diagnostic. A ``Case 2''-like system also peaks earlier at $t_{\rm {peak}} \approx \unit[4]{years}$ cf. $t_{\rm {peak}} \approx \unit[10]{years}$ for a ``Case 1''-like or a system without an engine.
Both these features are direct consequences of an afterglow brightness or the peak time primarily determined by the ratio of the kinetic energy to the interstellar medium density~\citep{sari99}, which are much higher in a ``Case 2''-like system. 
2) A ``Case 1''-like kilonova afterglow is brighter but difficult to distinguish (especially after peak) from the kilonova afterglow without an engine in both X-rays and radio.
This is especially true in light of the order of magnitude uncertainty in $\nu L_{\nu}$ caused by just the uncertainty in the interstellar medium density. A similar if not greater uncertainty exists in other afterglow microphysical parameters such as the fraction of accelerated electrons and the fraction of energy in the electrons and magnetic field that make decoupling these two kilonova afterglows virtually impossible. This difficulty is true even in a scenario like GRB170817A, where the jet afterglow is well observed and provides reasonable constraints on these parameters. 
3) The uncertainty in the kilonova afterglow parameters make it difficult to decouple the three representative cases after $\approx \unit[10-12]{years}$, with the difference in brightness most significant at early times, precisely when other processes such as the afterglow of the relativistic jet, the emergence of the magnetar wind nebula, or fall back accretion onto the black hole may muddy the picture~\citep{Hajela2021, Metzger2021}. 

\section{Inferring the presence of a neutron star engine}\label{sec:diagnostics}
Our current understanding of the neutron star mass distribution and the nuclear equation of state suggests that a significant fraction of binary neutron star mergers may produce a long-lived neutron star central engine~\citep{Margalit2019, Sarin2022}. 
This naturally leads to the question; how can we diagnose whether a given kilonova observation is a magnetar-driven kilonova or an ordinary one? 
Answering this question has significant ramifications on our understanding of the binary neutron star mass distribution and into inferences from the kilonova observations such as the Hubble constant~\citeg{Coughlin2020} or nuclear equation of state~\citeg{Margalit2017, Coughlin2019, Capano+20, Raaijmakers2021_eos, Nicholl2021}. 
Ultimately, determining the presence of a neutron star relies on understanding the systematic uncertainty in kilonova models and combining multiple different diagnostics.

\subsection{Kilonova model systematics}
In Fig.~\ref{fig:photometry}, we plotted the $r$, $g$, and $z$-band lightcurves for the representative cases described in Sec.~\ref{sec:cases}. In the same figure, the dashed curves represent the same representative case but with a different kilonova model; a kilonova model from~\citet{Metzger2019} including the same physics of gamma-ray leakage and spin-down evolution as described in Sec.~\ref{sec:model}. 
The only significant difference between the two models is how the mass is distributed, but this change alone can create significant differences. 
The model described in Sec.~\ref{sec:model} is a one-zone model that can be thought of as a single shell of mass with initial velocity $v_{\rm ej}$. In contrast, the model motivated by~\citet{Metzger2019} assumes a series of mass shells with total mass $M_{\rm ej}$ with an mass-averaged initial velocity $v_{\rm ej}$ assuming homologous expansion, i.e., faster matter with less mass is ahead of slower matter with more mass, with only the bottom layer being heated up by the neutron star. 
As Fig.~\ref{fig:photometry} demonstrates, this one change has a significant impact on the light curve, in particular on ``Case 2'', the representative case where the bulk of the energy goes into accelerating the ejecta, with the difference in ``Case 1'' not being as significant. This is consistent with intuition, as the dynamical evolution becomes more critical in ``Case 2'' as more energy accelerates the ejecta than in ``Case 1'' where the internal energy evolution plays a more prominent role. 

Observations of AT2017gfo are best fit with multiple components corresponding to different ejecta sources, such as one component for the dynamical ejecta and another for the ejecta from disk wind. 
In these works, both components are modelled analytically and separately~\citep{Smartt2017, Villar2017, Nicholl2021} ignoring the coupling between the components. With only one definitive observation of a kilonova, it is not clear whether multiple components are necessary to explain the observations or whether one single component composed of shells with different masses, velocities, and perhaps opacity can better explain the observations~\citeg{Tanaka2017_kne, Waxman+18, Hotokezaka2020}. 

Another source of uncertainty is nuclear physics itself, which has been demonstrated previously to contribute to up to an order of magnitude uncertainty at peak~\citep{zhu20,Barnes+21}. Both these factors, alongside other issues explored by numerical simulations~\citeg{Kawaguchi2022, Wu2022}, the significance of neutron precursor emission~\citep{Metzger2015_precursor}, shock heated ejecta~\citep{Gottlieb+18,Piro&Kollmeier18}, neutrino-driven winds~\citep{Metzger2018_gw170817}, interaction with the jet~\citep{Klion2021, Nativi2021, Nativi2022}, and viewing angle dependencies~\citep{Klion2022} indicate systematic uncertainties, which until resolved suggest that the relative brightness of a kilonova alone may not a good diagnostic for distinguishing an engine-driven kilonova from an ordinary one. This is especially true for ``Case 2''-like systems unless they are observed quite early.  
\subsection{Diagnostics}
Perhaps the cleanest diagnostic available to infer the presence of a neutron star engine are gravitational waves from the nascent neutron star itself. Unfortunately, current detector sensitivities make it unlikely that such a signal would be observable to relevant distances, with most emission models predicting a horizon distance of $\sim \unit[1-2]{Mpc}$ for advanced LIGO at design sensitivity~\citep{Gao2017, Sarin2018, abbott19_gw170817_postmergerII, Lander2020, Sur2021}. Systematic uncertainties in kilonova modelling and nuclear physics also suggest that the brightness alone may not be a good diagnostic for inferring the presence of a neutron star engine. However, multi-wavelength observations of kilonovae and kilonova afterglows offer other diagnostics. 

In most scenarios, the cleanest electromagnetic diagnostic is the spectra itself. Long-lived neutron star remnants cool through neutrino radiation~\citep{shapiro83}, increasing the electron fraction and therefore reducing the opacity, $\kappa$ of the disk-wind ejecta~\citep{perego14, Metzger2014b, fernandez16, lippuner17}. 
Longer lifetimes imply more neutrino radiation and, therefore, result in significantly more electron-rich ejecta with less abundance of heavy $r$-process elements. 
This statement can be used to calibrate spectra and abundance ratios of certain $r$-process elements to simulations of the neutron star lifetime and provide inferences into the lifetime of a neutron star from the observed spectra. 
For example, numerical calculations suggest that a remnant lifetime longer than $\sim 300$ ms will make the electron fraction, $Y_{e} \gtrsim 0.25$ in the disk outflow ejecta~\citep{Metzger2014b, lippuner17} producing up to a maximum of $r$-process element with atomic mass $A \gtrsim 130$ cf. a maximum atomic mass of $A \sim 200$ for a system where no long-lived neutron star is formed~\citep{Metzger2019}. However, see other calculations, which suggest a longer lifetime of $\sim 1$s for a similar electron fraction~\citep{sekiguchi16, Kawaguchi2020}. Observations of certain spectral lines would then be a powerful diagnostic for the lifetime of the remnant.
For example, observation of Strontium lines~\citep{Watson2019} and spectral observations at multiple epochs, suggests that the remnant of GW170817 did not promptly collapse into a black hole or produce an infinitely stable neutron star~\citep{Gillanders2022}.
This stresses the need for kilonova candidates to be followed up with telescopes capable of making spectral measurements and for continuing to improve the calibration of spectra through detailed numerical radiative transfer simulations.

The increase in electron fraction also changes the opacity of the ejecta and, therefore, change the peak timescales and the kilonova colour evolution, making this a potential diagnostic. 
For example, lanthanide-rich ejecta has a gray opacity $\kappa \approx \unit[30]{cm^2~g^{-1}}$ cf. $\kappa \approx \unit[1]{cm^2~g^{-1}}$ for lanthanide-poor ejecta~\citeg{Metzger2019}. 
A system where a neutron star remnant survives for~$\sim \unit[1]{s}$ may result in disk-wind ejecta with $Y_{e} \gtrsim 0.25$ corresponding to an opacity $\kappa \sim \unit[3-5]{cm^2~g^{-1}}$~\citep{Metzger2014b, lippuner17}.
The bulk of the inferences into the kilonova observations of GW170817 were performed by assuming a fixed opacity for different components~\citeg{Villar2017, Nicholl2017}. 
Relaxing this in inferences may still allow constraints on $\kappa$ to distinguish opacities between $30$ and $1$, and therefore diagnose cases between a powerful infinitely stable engine or an ordinary kilonova without an engine. 
However, it is not clear that constraints on $\kappa$ would be sufficiently strong to disentangle more intermediate cases; when a neutron star only survives for a short amount of time.
We note that these mappings between electron fraction and opacity are typically independent of the temperature or calibrated to the temperature when the light curve peaks~\citep{Tanaka2020} and may not be appropriate for the first few hours when magnetar spin-down energy is most relevant. 
Other features such as shock-heating, jet-interaction or viewing angle dependence may also muddy the waters~\citep{Klion2021, Nativi2021, Klion2022} suggesting that the colour evolution alone may be an inconclusive diagnostic.

As we showed in Fig.~\ref{fig:gamma}, across the entire parameter space of different magnetic field strengths, the presence of the neutron star increases the ejecta velocity over the initial value. 
A measurement of the average bulk velocity from the spectra may therefore provide another diagnostic for inferring the presence of a neutron star central engine.
We note that it may be difficult (given the uncertainties in kilonova spectra), to decouple the increase in average bulk velocity due to the presence of the magnetar from systematic uncertainties due to the spectral modelling or from the intrinsic velocity distribution of the ejecta. 
Spectral fitting of AT2017gfo strongly suggests that there is a disjoint in the composition of the ejecta~\citep{Gillanders2022}. 
However, the observations can not decouple whether this disjoint is due to the stratification of the ejecta or due to there being two (or more) distinct ejecta components. If the former is true, the stratification naturally implies a significant variation in velocities (assuming homologous expansion) which may be difficult to disentangle from the velocity increase if there is a neutron star central engine. 

The increase in velocity will also have a significant impact on the kilonova afterglow. In particular, a ``Case 2''-like system (where most of the spin-down energy goes into accelerating the ejecta) produces a significantly brighter kilonova afterglow that also peaks on shorter timescales. 
Even accounting for the uncertainty due to the uncertain interstellar medium density (see the shaded bands in Fig.~\ref{fig:afterglowlcs}), such an afterglow would be distinguishable from a scenario where there is no engine. 
However, a ``Case 1''-like system (where more of the available electromagnetic spin-down energy is thermalised) may not be distinguishable from the kilonova afterglow without an engine, especially in light of uncertainties in other microphysical parameters that could hide the difference we see in Fig.~\ref{fig:afterglowlcs}. 
These uncertainties also have the potential to make it difficult to distinguish a ``Case 2''-like system as well unless the constraints on these microphysical parameters are relatively strong thanks to multi-wavelength observations or constraints from the relativistic jet afterglow. 
We note that the non-detections of kilonova afterglow have already been used to rule out the presence of highly energetic magnetar central engines in several short gamma-ray bursts~\citep{Metzger2014_grb,Horesh2016, Fong2016, Klose2019, Liu2020, schroeder20, Ricci2021}. 

A potentially rare but smoking gun diagnostic would be the direct detection of emission from the magnetar wind nebula (MWN). Radio emission ($\sim$ 1 GHz) will likely be emitted on the same timescale as the kilonova afterglow, but X-rays and millimetre wavelength emission should be detectable on much shorter timescales. X-rays from the MWN will have a comparable luminosity to the optical kilonova emission and are observable on a timescale of a few hours to days. However, this is contingent on the ejecta being completely ionised by the MWN~\citep{Metzger2014a}. 
Idealized photoionization calculations suggest that this would generally be the case~\citep{Margalit+18}.
However, if the ejecta is not completely ionised, X-rays will not be visible until the photoelectric absorption optical depth drops to $1$, which will take $\approx \unit[1-2]{months}$~\citep{Murase2015}. If a relativistic jet is launched, the X-ray afterglow produced will likely be dominant on either of these timescales. Given the potential conflict with the emission from the jet, unabsorbed millimetre wavelength emission from the MWN may be a better diagnostic. In superluminous supernovae, millimetre wavelength emission is predicted to peak on timescales of $\sim 1$ year~\citep{Omand2018} and may peak earlier in a magnetar-driven kilonova due to the higher expected remnant magnetic fields, lower spin-down time, and lower ejecta mass. On this timescale, a relativistic jet could only produce a brighter afterglow if the observer was located significantly off-axis. We note that a significant uncertainty in these predictions is the unknown emission spectrum of the MWN, as the only known spectral models are calibrated to Galactic PWN \citep{Tanaka2010, Tanaka2013} and studies of SLSNe have only provided weak constraints or inconclusive results \citep{Law2019, Mondal2020, Eftekhari2021, Murase2021}. If observations of magnetar wind nebula are to become a powerful diagnostic, this emission spectrum needs to be studied in more detail. 

Although many of these diagnostics alone have potential issues that make their overall distinguishing power inconclusive, combining multiple diagnostics (especially ones that correspond to features of the spectra or the velocity, such as the kilonova and kilonova afterglow) offers the best prospects for distinguishing an ordinary kilonova from a kilonova with a powerful, infinitely stable engine. 
However, some problems may arise when considering the most likely outcome of a binary neutron star merger (given our current understanding of the binary neutron star mass distribution), a neutron star engine that only survives for a short period. 
Analytic calculations suggest that neutron stars born with a mass between $\Mtov{}$ and $1.2\Mtov{}$ can support themselves against gravitational collapse through rigid body rotation for up to $\unit[10^4]{s}$, assuming the neutron star only spins down due to vacuum dipole radiation~\citep{ravi14}. 
More realistic calculations incorporating a combination of electromagnetic and gravitational-wave spin down indicate that such neutron stars will collapse on much shorter timescales~\citep{fan13, gao16}. 
Assuming the time of the sharp drop in luminosity seen in short gamma-ray burst X-ray afterglows is due to the collapse of a neutron star, most collapse time measurements are $t_{\rm collapse} \lesssim \unit[300]{s}$~\citep{rowlinson13, Sarin2020a}. 
The collapse into a black hole reduces the rotational energy available in the system making them less distinguishable from an ordinary kilonova without an engine. However, a substantial amount of rotational energy must be reduced from the system before it becomes indistinguishable from a $\unit[10^{51}]{erg}$ ordinary kilonova. 
\section{Discussion}\label{sec:implications}
A significant fraction of binary neutron star mergers may result in the temporary formation of a neutron star central engine. 
This paper explored the diversity of kilonova and kilonova afterglows expected from such engine-driven explosions. We first explored the diversity due to changes in the internal and external magnetic field strengths and later in other properties such as the initial spin period, fraction of energy lost in pair cascades, and the kilonova modelling itself.
The biggest source of diversity originates from the remnant's internal and external magnetic field strengths, which dictates what fraction of rotational energy is available in electromagnetic radiation or lost in gravitational waves and what fraction of the available electromagnetic energy is thermalised or lost in accelerating the ejecta. 
Through two representative examples, we illustrated that a magnetar-driven kilonova is either ``bright'' (Case 1) or ``fast'' (Case 2), where either the kilonova itself is significantly brighter than the kilonova of a system without an engine (Case 1), or the kilonova afterglow is brighter and peaks on a shorter timescale (Case 2). 
Although we note that there exists a continuum between these two representative cases.
As kilonovae are fast transients, this stresses the need for low cadence surveys or low latency/negative latency gravitational-wave alerts to catch kilonovae before or at their peak, where the differences between an ordinary kilonova and a kilonova with an engine are at their largest. 

Even in the parameter space where gravitational-wave losses from a magnetic-field induced ellipticity are significant (Case 1), the engine-driven kilonova will be significantly brighter than an ordinary kilonova without an engine. This is a direct consequence of the fact that, in the region of parameter space where magnetic-field induced ellipticity is large, and gravitational-wave losses dominate occur,  more of the available electromagnetic energy is thermalised. In effect, this can be flipped to state that there is no realistic magnetar engine with significant gravitational-wave emission (from a magnetic-field induced ellipticity) where the signature of an engine could be effectively hidden away from the kilonova. In other words, even if a significant amount of energy is lost in gravitational waves, such engine-driven kilonovae will still be noticeably bright. 
This is always true for an engine that only radiates gravitational waves through a magnetic deformation unless $f_{\rm EM} \lesssim 10^{-3}E_{\rm rot}$, which is only possible for $B_{\rm ext} \lesssim \unit[10^{15}]{G}$ and ellipticity $\epsilon_B \approx 10^{-2}$ which is not realistic. 
By contrast, a ``Case 2''-like system with no significant gravitational-wave radiation can produce a kilonova that is difficult to distinguish from an ordinary one. However, the kilonova afterglow becomes a powerful probe here as it will be brighter and peak earlier. We note that a ``Case 1''-like kilonova afterglow is indistinguishable from the kilonova afterglow of an ordinary kilonova, given typical uncertainties in afterglow parameters. This emphasises the need for combining diagnostics that can probe the opacity/elements synthesised and the kinetic energy/velocity to infer the presence of a neutron star. 
Our finding that a powerful long-lived post-merger magnetar remnant cannot be ``hidden'' even if gravitational-wave spindown is important (i.e. it will distinctly manifest in the observable kilonova and/or kilonova afterglow) strengthens previous arguments that GW170817 did not form such a remnant, and has important implications for the nuclear equation-of-state~\citeg{Margalit2017}.

Throughout this work, we assumed that the only gravitational-wave emission mechanism is a magnetic-field induced deformation. 
However, other channels, such as the secular bar-mode or r-mode instability, are also potentially active in a newly born neutron star~\citep{corsi09, andersson01}. 
For example, \citet{doneva15} find that the secular bar-mode instability has a maximum saturation amplitude of $10^{-6}Mc^2$. For the neutron star parameters we used above, this translates into an effective bar-mode ellipticity of $2 \times 10^{-3}$~\citep{doneva15, lasky16}. 
If the bar-mode instability is active, it could make it more difficult to distinguish an engine-driven kilonova to one without an engine by reducing the available electromagnetic radiation energy across the entire parameter space. However, it is difficult to predict the bar-mode saturation timescale and ultimately how the bar-mode torque competes with the electromagnetic torque.
Another energy-loss channel is a relativistic jet, which (if launched) must be powered by the available electromagnetic spin-down energy.
However, given that typical short gamma-ray burst jet energies are only $\sim \unit[10^{50}]{erg}$, this is likely not difficult to accommodate in the large rotational energy reservoir, apart from situations where gravitational-wave losses are significant. 
Across the entire parameter space, to effectively hide the signature of the magnetar from both the kilonova and kilonova afterglow one needs $f_{\rm EM} \lesssim 10^{-3}-10^{-2} E_{\rm rot}$ for the most powerful neutron star engines. However, in reality, these fractions are much higher, particularly if only one diagnostic is used, until several systematics are resolved.

Throughout our analysis, we assumed $\chi = \pi/2$, i.e., that the neutron star is an orthogonal rotator and remains in that state. However, this is not wholly accurate. Numerical simulations show that (depending on the initial angle and other parameters) a newly born neutron star takes some small fraction of time ($\sim$ seconds to minutes) to become orthogonal, remains an orthogonal rotator for $\sim \unit[10^{6}]{s}$, then slowly becomes an aligned rotator over a timescale of tens to hundreds of years; see Fig.~4 of \citet{Lander2020}. The rotational energy lost through gravitational-wave emission is most efficient when the system is an orthogonal rotator. 
Not modelling the very early evolution implies we are being conservative in the overall fraction of rotational energy available to affect the kilonova. 

The same effect applies when the neutron star starts to become an aligned rotator again. In the model of \citet{Lander2020}, this happens when the free precession timescale of the star becomes equal to the time scale of the particle reactions that tend to restore beta equilibrium in the stellar matter.  \citet{Lander2020} find that this happens at $\approx \unit[10^{6}]{s}$ for a neutron star with the same parameters as ``Case 2'' described earlier (again, see their Fig.~4), by which time there is relatively little rotational energy left, so that the kilonova and kilonova afterglow properties are unaffected.  Note, however, that this timescale is sensitive to the microphysics of these weak interactions, and also to the internal temperature, so could be shorter.  

As noted in Sec.~\ref{sec:diagnostics}, the emission spectrum of a newborn magnetar wind nebula (MWN) is not well constrained. However~\citet{Ai2022} present a model for the shock structure of a magnetar-driven kilonova that has interesting implications on the MWN spectrum. 
Their model includes a Poynting-flux dominated magnetar wind which generates
a reverse (termination) shock upon collision with the kilonova that propagates back to the neutron star at $\sim c$ and dissipates when reaching the wind-launching region on a timescale of a few seconds. 
The unshocked wind that reaches the kilonova and eventually leaks out will result in a spectrum significantly different from a regular pulsar wind nebula (PWN) spectrum, which is typically from a shocked wind~\citep{Gaensler2006}. 
In a typical PWN, the reverse shock accelerates electrons and positrons to ultra-relativistic velocities, where they can produce broadband synchrotron radiation and produce gamma-rays through inverse Compton scattering~\citep{Ginzburg1965, Gaensler2006}. However, with a highly magnetized wind and no reverse shock, there would only be low energy cyclotron/synchrotron radiation.
There are some ways to avoid this scenario. 
If the wind is electron-positron pair-dominated instead of Poynting-flux dominated or the bulk velocity of the wind is high, then the reverse shock can be steady and not dissipate. Alternatively, magnetic reconnection in a current sheet generated by a striped wind~\citep{Lyubarsky2001, Sironi2011} or a kink instability in the field~\citep{Begelman1998, Porth2013} may also alleviate this issue. 
The reverse shock may also be able to reform once the neutron star has lost a significant amount of energy or the kilonova has expanded a significant distance.

The disappearance of the reverse shock may also have implications for other magnetar-driven transients. In the magnetar model for superluminous supernovae (SLSNe), the shocked wind is absorbed by the ejecta and powers the optical light curve, just as in magnetar-driven kilonovae.  It's unknown whether an unshocked wind could generate the same supernova luminosity, or if the previously detected radio counterparts~\citep{Law2019, Eftekhari2019, Coppejans2021b, Coppejans2021a} could be consistent with this model.

Our model and some of our results are also applicable more generally to other transients, such as SLSNe and some broad-line Type Ic supernovae (BL-Ic SNe), both of which could have a magnetar-driven sub-population.
BL-Ic supernovae are predicted to be powered by neutron stars with $B_{\rm ext} \gtrsim 10^{14.5}$ G and have ejecta masses $\sim$ 1 $M_\odot$, while SLSNe are predicted to be powered by neutron stars with $B_{\rm ext} \approx 10^{13-14.5}$ G and have larger ejecta masses, sometimes up to $\sim$ 30 $M_\odot$~\citep{Nicholl2017, Suzuki2021, Chen2022}.  
Since these transients have similar magnetar engines to our kilonovae, they could also lose a significant amount of energy to gravitational radiation~\citep{Kashiyama2016} and have ejecta accelerated by their MWN~\citep[e.g.][]{Murase2015, Suzuki2021}.  
Current models widely used in inference~\citep{Nicholl2017} do not capture these nuances, instead assuming the ejecta velocity to be a free parameter that does not vary with time and is completely decoupled from the magnetar luminosity. However, as shown in~\citep{Suzuki2021} and in Fig~\ref{fig:calor}, a significant fraction of the neutron star spin-down luminosity is converted into kinetic energy (depending on the ratio of $t_{\rm SD}$ to $t_{\rm diff}$), making this assumption not-realistic for these transients.

The central magnetar is also expected to influence the nebular optical spectrum of the transient in each case, since the engine can photoionize the ejecta. For example, SN2012au, a suspected magnetar-driven supernova, was observed with an unexpectedly high amount of forbidden oxygen emission at $\sim$ 6 years post-explosion \citep{Milisavljevic2018}. However, there are not yet any reliable models for the nebular spectra of magnetar driven transients, although progress is being made~\citep[e.g.][]{Jerkstrand2017, Pognan2022b, Pognan2022a}. One interesting difference between magnetar-driven kilonovae and supernovae is the appearance and role of dust.  In supernovae,  dust can form in the ejecta after $\sim$ a few months, and when there is a central magnetar present, the dust can be heated by absorbing radiation from the MWN and re-radiate that energy in the infrared \citep{Omand2019, Chen2021}.  However, conditions in kilonovae, such as the generally higher ejecta temperature, velocity, and density may prevent formation of large clumps of dust.

With the multi-messenger era finally underway, improving our capabilities to effectively diagnose the engines of kilonovae has important ramifications on our understanding of the binary neutron star mass distribution and the nuclear equation of state. We have shown that differences in the remnant neutron star's internal and external magnetic field strength drives the majority of the diversity seen in magnetar-driven kilonovae. And combining diagnostics, in particular observations of the spectra, and kilonova afterglow together can distinguish magnetar driven kilonova across the bulk of the parameter space apart from circumstances where gravitational-wave emission from mechanisms other than a magnetic-field deformation is active and dominant or the neutron star collapses into a black hole shortly after formation. As our understanding of kilonova systematics improves, and observations become plentiful, magnetar-driven kilonovae could become a powerful and novel playground for understanding the early lives of nascent neutron stars. 


\section{Acknowledgments}
We are grateful to Shunke Ai and Paul Lasky for helpful discussions. We are also grateful to the anonymous reviewer for their comments on the manuscript. N.S is supported by a Nordita Fellowship, Nordita is supported in part by NordForsk. BM is supported by NASA through the NASA Hubble Fellowship grant \#HST- HF2-51412.001-A awarded by the Space Telescope Science Institute, which is operated by the Association of Universities for Research in Astronomy, Inc., for NASA, under contract NAS5-26555. DIJ acknowledges support from the STFC via grant number ST/R00045X/1.
\section{Code and Data Availability}
The kilonova models derived in this paper are implemented in the publicly available software for fitting electromagnetic transients,  {\sc{redback}}~\citep{git_repository}. No data was used in this manuscript. 

\bibliographystyle{mnras}
\bibliography{ref}


\bsp    
\label{lastpage}
\end{document}